\documentclass{article}

\usepackage{graphicx}
\usepackage{caption}
\usepackage{subfig}
\usepackage{amsmath}
\usepackage[english]{babel}
\usepackage[autostyle]{csquotes}
 \usepackage{mathptmx}      
\usepackage{latexsym}
\begin{document}

\noindent	
{\large{\bf A step function density profile model for the convective stability of $CO_2$ geological sequestration}}\\
\vspace{0.10in}	
	
\begin{center}	
    C. Taber Wanstall{\footnote{ctwanstall@crimson.ua.edu}}\\  and Layachi Hadji{\footnote{lhadji@ua.edu (corresponding author)}}\\
		The University of Alabama, 	Tuscaloosa, Al 35487-0350\\
	\end{center}	
\vspace{0.1in}	

\section{Abstract}
The convective stability associated with carbon sequestration is usually investigated
by adopting an unsteady diffusive basic profile to account for the space
and time development of the carbon saturated boundary layer instability. The method of normal modes is not applicable due to the time dependence of the nonlinear base profile. Therefore, the instability is quantified either  in terms of critical times at which the boundary layer instability sets in or in terms of long time evolution of initial disturbances. This paper adopts an unstably stratified basic profile having a step function density with top heavy carbon saturated layer  (boundary layer) overlying a lighter carbon free layer (ambient brine). The resulting configuration resembles that of the Rayleigh-Taylor problem with buoyancy diffusion at the interface separating the two layers. The discontinuous reference state satisfies the governing system of equations and boundary conditions and pertains to an unstably stratified motionless state. Our model accounts for anisotropy in both diffusion and permeability and chemical reaction between the carbon dioxide rich brine and host mineralogy. We consider two cases for the boundary conditions, namely an impervious lower boundary with either a permeable (one-sided model) or poorly permeable upper boundary. These two cases posses neither steady nor unsteady unstably stratified equilibrium states. We proceed by supposing that the carbon dioxide that has accumulated below the top cap rock forms a layer of carbon saturated brine of some thickness that overlies a carbon-free brine layer. The resulting stratification remains stable until the thickness, and by the same token the density, of the carbon saturated layer is sufficient to induce the fluid to overturn. The existence of a finite threshold value for the thickness is due to the stabilizing influence of buoyancy diffusion at the interface between the two layers. With this formulation for the reference state, the stability calculations will be in terms of critical boundary layer thickness instead of critical times, although the two formulations are homologous. This approach is tractable by the classical normal mode analysis. Even though it yields only conservative threshold instability conditions, it offers the advantage for an analytically tractable study that puts forth expressions for the carbon concentration convective flux at the interface and explores the flow patterns through both linear and weakly nonlinear analyses.\\


\section{Introduction}
\label{intro}

A growing concern in today's fossil fuel based society of energy production and consumption
is the effect combustion byproducts have on the environment; more specifically
the greenhouse effects of carbon dioxide \cite{1}-\cite{3}. Geological $CO_2$ sequestration
has been found to be one of the most effective short-term solutions to alleviate $CO_2$
greenhouse effects. Some innovative strategies have recently been proposed that reduce
the risk of the carbon dioxide leaking back into the atmosphere \cite{4}. From a
modeling standpoint, one of the key objectives is to quantify the mechanisms and
time scales associated with the possible migration of $CO_2$ back into man's environment.
We refer the reader to the recent papers \cite{5,6} and references therein for more details.
Quantitative studies of $CO_2$ dissolution into saline aquifers are of great importance
to help predict long term effects. Carbon dioxide dissolution rates into brine aquifers
are controlled by three main mechanisms: 1) diffusion of $CO_2$ out of the supercritical
phase into the brine as well as the diffusion within the brine, 2) chemical reactions
between the $CO_2$ rich brine and host mineralogy, and 3) natural convection driven
by the density gradients between the $CO_2$ saturated and unsaturated brine \cite{7}. A stability
analysis will be preformed to quantify the point at which natural convection
will occur. The mathematical model accounts for anisotropy in both diffusion and
permeability, and chemical reactions between the carbon dioxide present in the brine
with the host mineralogy. In this study the chemical reactions will be modeled as first
order, the porous medium will be anisotropic with constant vertical and planar diffusion
coefficients, and a permeability that slowly decreases with depth.\\
\\

Much of our understanding of buoyancy-driven instabilities originated from the
study of Rayleigh-B\'{e}nard convection wherein a horizontal fluid layer is confined
between two plates of infinite horizontal extent, the lower of which is heated. An
unstable density stratification develops and causes the fluid to overturn, provided the
heating level exceeds a certain threshold value. There are other instances where the
density stratification is caused by internal heating, in which case the unstable stratification
occurs only over a small portion of the fluid layer. For example, Batchelor and
Nitsche \cite{8} carried out a theoretical investigation wherein the fluid is unstably stratified
only over a thin centrally located region. They obtained the threshold conditions
for instability as function of the thickness of the unstable layer and also considered the
implications on the instability as the thickness approaches zero. Simitev and Busse \cite{9} (SB) studied a model of convection with a heat source distribution described by $Q(z) = 2 \gamma \tanh{(\chi z)} /(\cosh^2{(\chi z)}$. The resulting static temperature distribution is then given by $T_B(z) = b z - \tanh{(\chi z)}/\chi$. The constants $b$ and $\chi$ are measures of stable and unstable stratification, respectively, with the thickness of the unstably stratified layer getting thinner as $\chi$ increases resulting in a situation that mimics B\'{e}nard convection without the presence of the horizontal boundaries. Hadji {\it et al.} \cite{10} considered the case of a step change profile which would correspond to the SB profile in the limits of both $b \to 0$ and $\chi \to \infty$, namely, $T_B(z) = T_1 + (T_2 - T_1) {\cal H}(z-Z_0)$, where $T_1$ and $T_2$ are the temperature values at the lower and upper plates, respectively, $T_1>T_2$ and ${\cal H}(z)$ denotes the Heaviside function. Such a profile results in a heavy layer on top of a lighter one separated by a horizontal interface $z=Z_0$ across which diffusion of buoyancy takes place. This profile leads to instability threshold conditions that are much lower than those corresponding to a linear continuous unstable stratification.\\

The super-critical $CO_2$ that is injected into the denser brine formation ends up being sandwiched
between an impervious top cap rock and the lower ambient brine in the
saline aquifer with a gas-brine interface emerging between the two layers. The carbon dioxide that is right at the gas-brine interface then diffuses and dissolves slowly into the brine, the density of which increases relative to the unsaturated brine and triggers
the development of an unstable stratification \cite{7}. The general scenario that has emerged from experiments is that a quasi-stable regime is initially attained with the formation of a boundary layer that subsequently grows and becomes unstable to convective overturning when some critical thickness is reached. At this point, the boundary layer has acquired enough potential energy to overcome the stabilizing effect of buoyancy diffusion \cite{11}-\cite{14}. From the mathematical standpoint, the motionless state is governed by the diffusion equation with a Dirichlet and Neumann boundary conditions for the concentration at the top and bottom walls, respectively, and subject to some initial value. The solution decays to a constant value as time goes to infinity (cf. Eq. (21) in \cite{7}) thus precluding the existence of an unstably stratified equilibrium base state.
Several lines of inquiry have been pursued to theoretically quantify and determine the threshold instability conditions. These studies differ mainly in the way they have adopted and approximated the basic equilibrium density profile, scaled the governing equations and formulated the boundary conditions. Some investigations have adopted a linear steady diffusive basic state profile to model the carbon sequestration as a typical convection problem in a fluid saturated medium (e.g. \cite{15}-\cite{20}). These studies have considered Dirichlet boundary conditions at both top and lower walls to obtain an unstably stratified steady equilibrium reference state, the stability of which can then be carried out.
The nonlinear behavior of the steady profile considered in \cite{15} for the one-sided model is due solely to the presence of the reaction term in the model, without which the equilibrium state is uniform and thus stably stratified. Other investigations have considered approximations to the unsteady pure diffusion boundary-layer profile
 (cf. \cite{7}, \cite{21}-\cite{23}) and references therein) to take into account the development of the carbon saturated boundary layer and its instability. The approximations varied from truncated infinite series to similarity type solutions when the finite depth is asymptotically extended to infinity. Due to the time dependence of the nonlinear base profile, the instability threshold conditions are then expressed in terms of either critical times at which the boundary layer instability sets in \cite{7} or in terms of growth rates of the critical modes corresponding to the most dangerous disturbance \cite{23}-\cite{24}. The critical time for instability corresponds to a boundary layer having reached its critical thickness so that it is prone to convective overturning.\\
 
In this paper, we consider two cases for the boundary conditions, namely an impervious lower boundary with either a permeable or poorly permeable upper boundary. These two cases posses neither steady nor unsteady unstably stratified equilibrium state. We proceed by supposing that after some finite time has elapsed, the carbon dioxide that has accumulated at the top forms a layer of carbon saturated brine of some thickness that overlies a carbon-free brine layer. The resulting stratification remains stable until the thickness, and by the same token the density, of the carbon saturated layer is sufficient to induce the fluid to overturn. The existence of a finite threshold value for the thickness is due to the stabilizing influence of buoyancy diffusion at the interface between the two layers. Thus, we consider the situation wherein the lighter brine-free layer of thickness $(Z_0d)$, $0<Z_0<1$, underlies a heavier carbon saturated brine of thickness $(d-d\,Z_0)$ and concentration $C_B$.  The definition of the density distribution $C_B(z)$ depends on the degree of approximation. For instance, a linear piece-wise approximation will have a reference concentration of the from,
$C_{ref}(z)=(z-Z_))/(1-Z_0)$ for $Z_0<z<1$ and $0$ elsewhere. With this formulation for the reference state, the stability calculations will be in terms of critical boundary layer thickness, $Z_0$, instead of critical times, although the two formulations are homologous.
Thus, this approach has the advantage of being tractable by the classical normal mode analysis. In the foregoing analysis, we consider the simple case of a reference state consisting of a unit step function that mimics the Rayleigh-Taylor instability. The resulting basic stratification is the most unstable and is expected to yield conservative values for the instability threshold conditions \cite{10}. It offers, however, the advantage of conducting an analytically tractable investigation to explore the flow patterns through both linear and weakly nonlinear analyses.\\

The plan for the remainder of the paper is stated as follows. Except for the dependence of the permeability on depth, our mathematical formulation in Sec. 2 is similar to Hill and Morad's model \cite{15}. For this model, the exchange of instabilities has been shown to hold. In Sec. 3, the linear stability analysis is carried out for the case of a permeable top boundary and for two cases, namely a constant permeability and a permeability that increases with depth. In Sec. 3, we will also consider the case of nearly-impermeable top boundary and both constant and variable permeability. The constant permeability case from Sec. 3 is extended to the weakly nonlinear analysis in Sec. 4 where a nonlinear evolution equation for the concentration is derived
and a uniformly valid super-critical steady state solution is put forth. A discussion and concluding remarks are presented in Sec. 5.
\section{Formulation}
\begin{figure}[h]
			\centering
			\includegraphics[scale=0.5]{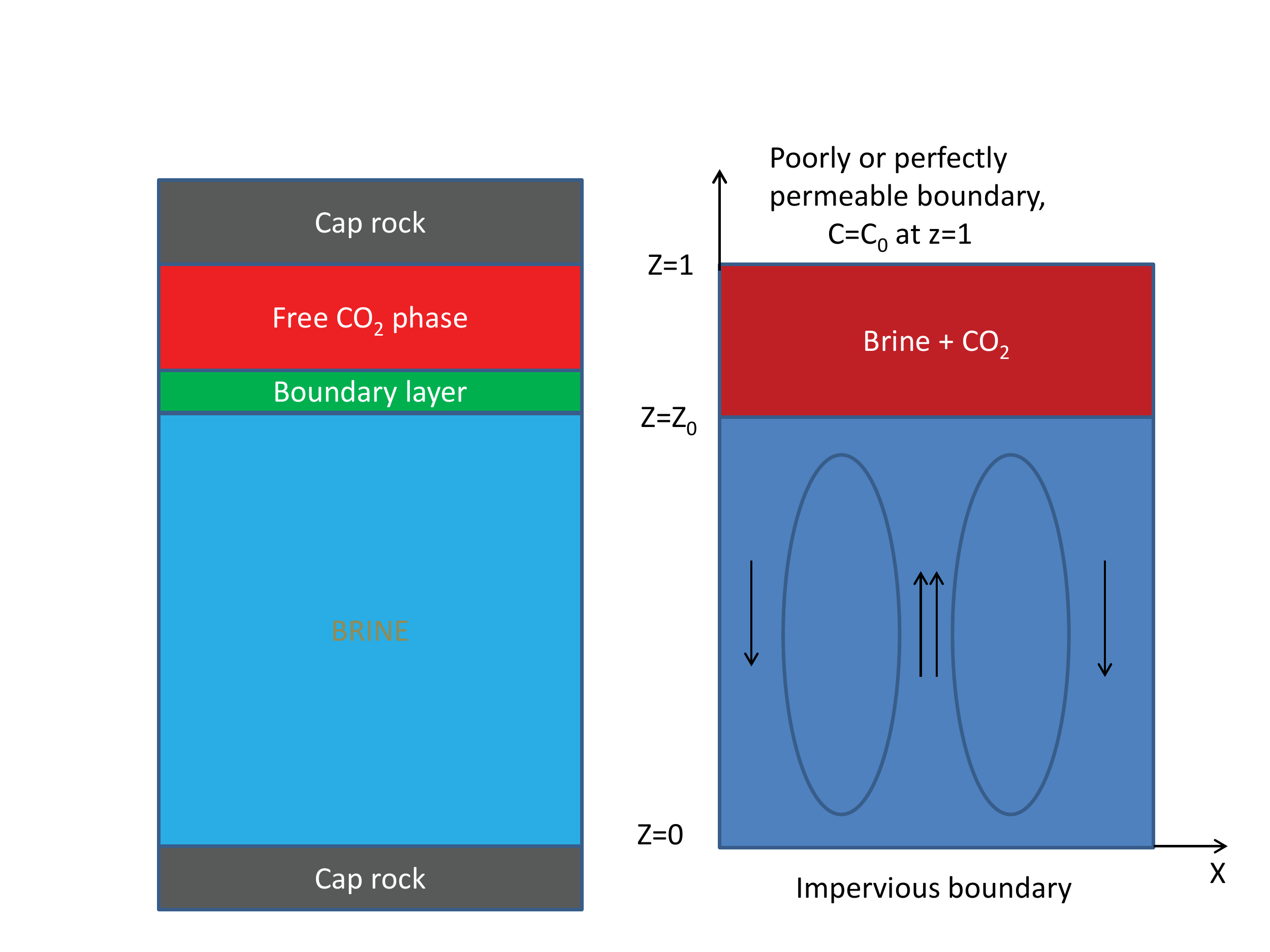}\label{math}
			\caption{A sketch of a $CO_2$-rich brine layer overlying a carbon-free layer. The two layers are confined between two plates located at $z=0$ and $z=1$ and separated by a horizontal interface at $z=Z_0$. (Left) A schematic of the experimental evidence carried out by Neufeld {\it et al.} \cite{14}. (Right) Simplified model of the experimental evidence.}
			\label{fig:1}
		\end{figure}
We consider a fluid saturated porous medium of infinite horizontal extent that is confined between two plates that are a distance 
$d$ apart. The model accounts qualitatively for some non-uniformities in the rock formation by incorporating the space variations of both diffusion and permeability.  Thus, we account for some contrast between the vertical and horizontal diffusion of the carbon dioxide by considering a diffusivity tensor of the form ${\cal K} = {\bf I}\,<\kappa_h, \kappa_v>$, where ${\bf I}$ is the $2 \times 2$ identity matrix.  Most formations have a permeability that decreases with depth which we model its space variations  by  assuming that  $K=K_0\,\Pi(z)$, where $\Pi(z)$ is a decaying exponential function of depth. There are other models of formation permeability that were considered within the context of the problem of carbon dioxide sequestration and associated convection. For instance, Paoli {\it et al.} \cite{6} introduced the ratio of lateral to vertical permeabilities to account for the space variations of permeability and carried out a numerical study on the influence of this ratio on the flow patterns. It is noteworthy to observe that, due to the way the equations were scaled in (cf. Eq. (12) in \cite{6}), this ratio happens to play the same role as our ratio of diffusivities $\xi$. Hill and Morad \cite{15} considered a linear function of depth which we retrieve with our exponential model when we consider a slowly decaying exponential function. It is also noteworthy to observe that these models not only capture qualitatively the space variation of permeability and diffusivity, but their simple mathematical form allows for the derivation of purely analytical results. With the assumptions that both Darcy's law and the Boussinesq approximation hold, the governing system of non-dimensionalized equations, which consists of Darcy's equation, the conservation of mass and carbon dioxide equations and an equation of state, is described by \cite{15}
\begin{equation}\label{epmass}
{\bf \nabla} \cdot \bf{u}=0
\end{equation}
\begin{equation}\label{epdarcy}
\frac{\bf{u}}{\Pi(z)}=-{\bf \nabla} p -c\,\bf{k}  
\end{equation}
\begin{equation}\label{epspecies}
\frac{\partial c}{\partial{t}}+ {\bf u} \cdot {\bf \nabla}\,c  + {(dM/dz)}\,w={(\xi}/{R}){\bf \nabla}_h^2\,c+({1}/{R}) \frac{\partial^2 c}{\partial z^2}-(Da/R)\,c 
\end{equation}
\begin{equation}
\rho=\rho_0[1+q_c(c-C_0)]
\end{equation}

where ${\bf u}$ is the fluid velocity, $c$ is the deviation of the concentration in volume fraction from the diffusive state, $M(z)$ is the basic concentration profile, $p$ is the pressure, ${K_0\Pi(z)}$ is the depth-dependent permeability, where $K_0$ is some reference permeability value and ${\bf k}$ is the unit vector that is in opposite direction of the gravity vector, ${\bf g} = - g{\bf k}$. The dimensionless parameter $R$ is the Rayleigh or Darcy-Rayleigh number $R = q_C\,g\,d\,K_0\, C_0/\varphi_p\nu\,\kappa_v$ where $q_C$, $C_0$, $\nu$, $\kappa_v$, $\rho_0$ and $\varphi_p$ are the coefficient of solutal expansion, a reference concentration of $CO_2$ taken to be the concentration right at the top of the brine layer, the kinematic viscosity, the vertical $CO_2$ diffusion coefficient, a reference density and the porosity,  respectively. The other parameters that appear in Eq. (\ref{epspecies}) are the ratio of horizontal and vertical solutal diffusivities $\xi = \kappa_h/\kappa_v$ and the Damk\"{o}hler number  $Da={\cal R}\,d^2/\varphi_p\,\kappa_v$, where 
${\cal R}$ is the reaction rate. As in \cite{15}, the above system was made dimensionless by using the scales $d$, 
$K_0\,g\,C_0\,q_c/\nu$, $\varphi_p\,d\,\nu/(K_0\,g\,C_0\,q_c)$ and $C_0$ for length, velocity, time and concentration, respectively. As discussed in the first section, the forgoing analysis pertains to a diffusion profile that consists of a heavy layer, $Z_0 < z<1$, on top of a lighter one, $0 < z < Z_0$, i.e., $M(z) = {\cal H}(z - Z_0)$, where ${\cal H}(z)$ is the Heaviside function.
Upon considering the vertical component of the double curl of Eq. (\ref{epdarcy}) and introducing the poloidal representation for the divergence free velocity field,
${\bf u} = \nabla \times (\nabla \times \phi\,{\bf k})$, the above system of equations reduces to
\begin{eqnarray}
\label{system2}
f\,\nabla^2 \phi - (f^\prime)\,\phi_z&=&-f^2\, c\\
\label{system3}
{c_t} + \phi_{xz}\,c_x + \phi_{xx}\,c_z &=& -\nabla^2_h \phi\,\delta (z-Z_0)+(\xi\,{\nabla_h}^2 c+{c_{zz}}- {Da}\,c)/R
\end{eqnarray}
 where $\delta(z)$ is the Dirac delta function. The function $f(z)$ results from the transformation of $\Pi(z)$, namely $f(z)=\Pi(z/d)$
 and models the variation of the permeability with depth with $f(z):=1$ when there is no change in permeability. Both
 subscript and prime notations refer to differentiation. Two models will be
 presented in this paper. The first is a perfectly permeable upper boundary described by
 $c = 0$ at $z = 1$ with an impermeable lower boundary which has been the focus of
 most investigations and generally labeled as the one-sided model \cite{6}. The second is a top bounding plate 
 that is assumed to be rigid
 and nearly-impermeable to carbon dioxide leak while the lower boundary is assumed
 rigid and perfectly impervious. These assumptions lead to Dirichlet condition for the
 velocity, Neumann condition for c at the lower boundary and either a Dirichlet or a
 Newton law of cooling type condition for the concentration at the upper boundary,
 \begin{equation}
 \phi = 0, \quad {\mbox{at}} \quad  z=0, \quad {\mbox{and}} \quad z=1,
 \end{equation}
 \begin{equation}
 \begin{aligned}
 \label{bc2}
 &{\partial c \over \partial z} = 0 \quad {\mbox{at}} \quad z=0,\quad {\mbox{and}} \quad  
 \quad c=0, \quad z=1  \\
 & {\mbox{or}} \quad {\partial c \over \partial z} = 0,  \quad {\mbox{at}} \quad z=0 ,  \quad{\partial c \over \partial z} = -\beta\,c,  \quad {\mbox{at}} \quad z=1,
 \end{aligned}
 \end{equation}
where the mass transfer Biot number $\beta \ll 1$. The Newton law of cooling type boundary condition in Eq. (\ref{bc2}) assumes that  after the step function profile has been established, a small amount of carbon dioxide escapes at the upper boundary. This boundary condition allows for a nonlinear analysis to be tractable by long wavelength asymptotics. 

\section{Linear stability analysis}
\subsection{Constant permeability}

Upon linearizing Eqs. (\ref{system2},\ref{system3})  and introducing the Fourier modes, $< \phi, c> = <W(z),S(z)> e^{\imath\,\alpha\,x}$, we obtain the two equations, namely
\begin{equation}\label{stability1}
 \alpha^2 R\,W \delta(z-Z_0) 
= -\alpha^2\,{\xi}\,S+\,\frac{d^2 S}{dz^2}- Da\,S.
\end{equation}

\begin{equation}\label{stability2}
-\alpha^2\,f\,W+f\,\frac{d^2W}{d z^2} -f'\frac{dW}{dz}=-f^2 S
\end{equation}\\
with boundary conditions that correspond to the perfectly permeable upper boundary. The next step is to alleviate the $\delta$ function by splitting the equations up on the interval $0<z<Z_0$, and $Z_0<z<1$. Upon relaxing the function $f(z)$ to be constant and applying the substitution, $-p^2=-\xi\alpha^2-Da$, we solve Eqs. (\ref{stability1},\ref{stability2}) in the two separate regions, $0<z<Z_0$ and $Z_0<z<1$ which we label with the super scripted variables $W^-(z)$, $W^+(z)$, $S^-(z)$, and $S^+(z)$, respectively. With $D= d/dz$, we have
$$
(D^2-p^2){S^+}=0, \quad (D^2-\alpha^2){W^+}=-\alpha^2{S^+}
$$
\begin{equation}
\label{stab2}
(D^2-p^2){S^-}=0, \quad
(D^2-\alpha^2){W^-}=-\alpha^2{S^-}
\end{equation}
Upon solving the eighth-order system of differential equations described by Eqs. (\ref{stab2}), four of the constants of integration 
are determined by applying the boundary conditions at $z=0$ for variables super-scripted $(-)$ and at $z=1$ for the variables super-scripted $(+)$. For $f=1$, Eqs. (\ref{stability1}, \ref{stability2}) can be reduced to a single forth order ODE for $W$.
\begin{equation}
[D^2-(\alpha^2 \xi+Da)][D^2-\alpha^2]W=-\alpha^2 R W \delta(z-Z_0)
\label{4thode}
\end{equation} 
The four remaining constants are obtained by imposing continuity of $W$, $DW$ and $D^2W$ at $z=Z_0$; and the jump condition obtained by integrating Eq. \ref{4thode} over the interval $[Z_0-\ell,Z_0+\ell]$ and letting $\ell$ approach zero.
 The resulting fourth order homogeneous linear system of equations for the undetermined constants can now be expressed as ${\bf Q}\,{\bf Y} = {\bf 0}$ where the vector ${\bf Y} = <A^-\,A^+\,B^-\,B^+>$ and ${\bf Q}$ is a $4 \times 4$ matrix. An example of such a system is depicted in the Appendix for the case $\xi=1$ and $Da=0$ and for the boundary conditions described by Eqs. (\ref{bc2}). The homogeneous system
${\bf Q}\,{\bf Y} =0$ has a non-trivial solution if and only if $\det{(\bf Q)}=0$. When the latter is written as
$\det{(\bf Q)}= \det{(\bf Q_1)} - R \det{(\bf Q_2)}=0$, we obtain an expression for the Rayleigh number $R$, namely
$ R =  \det{(\bf Q_1)}/\det{(\bf Q_2)}$.\\
\\
\begin{figure}
	
		\centering
       \includegraphics[scale=0.42]{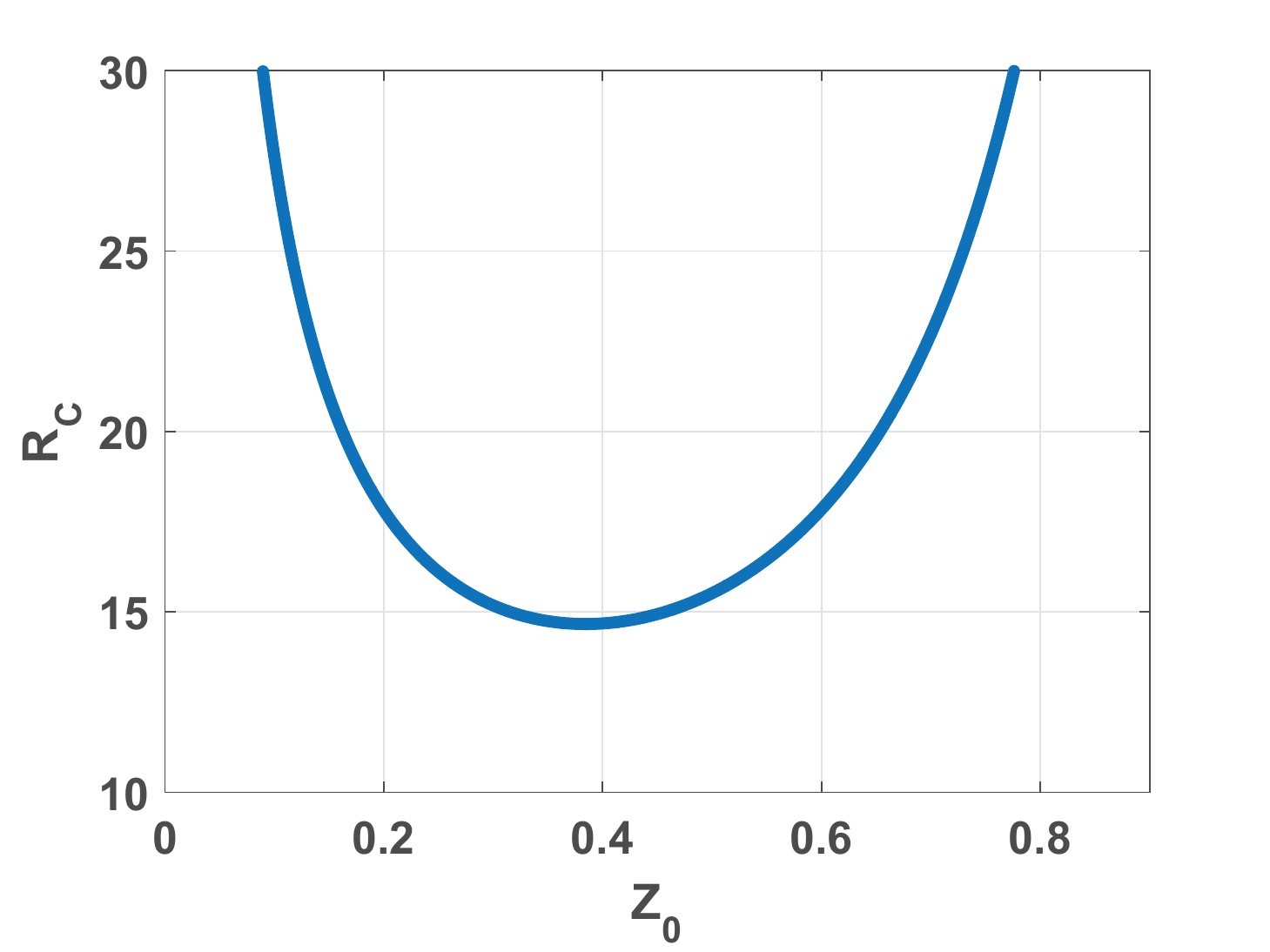}\\
       \includegraphics[scale=0.40]{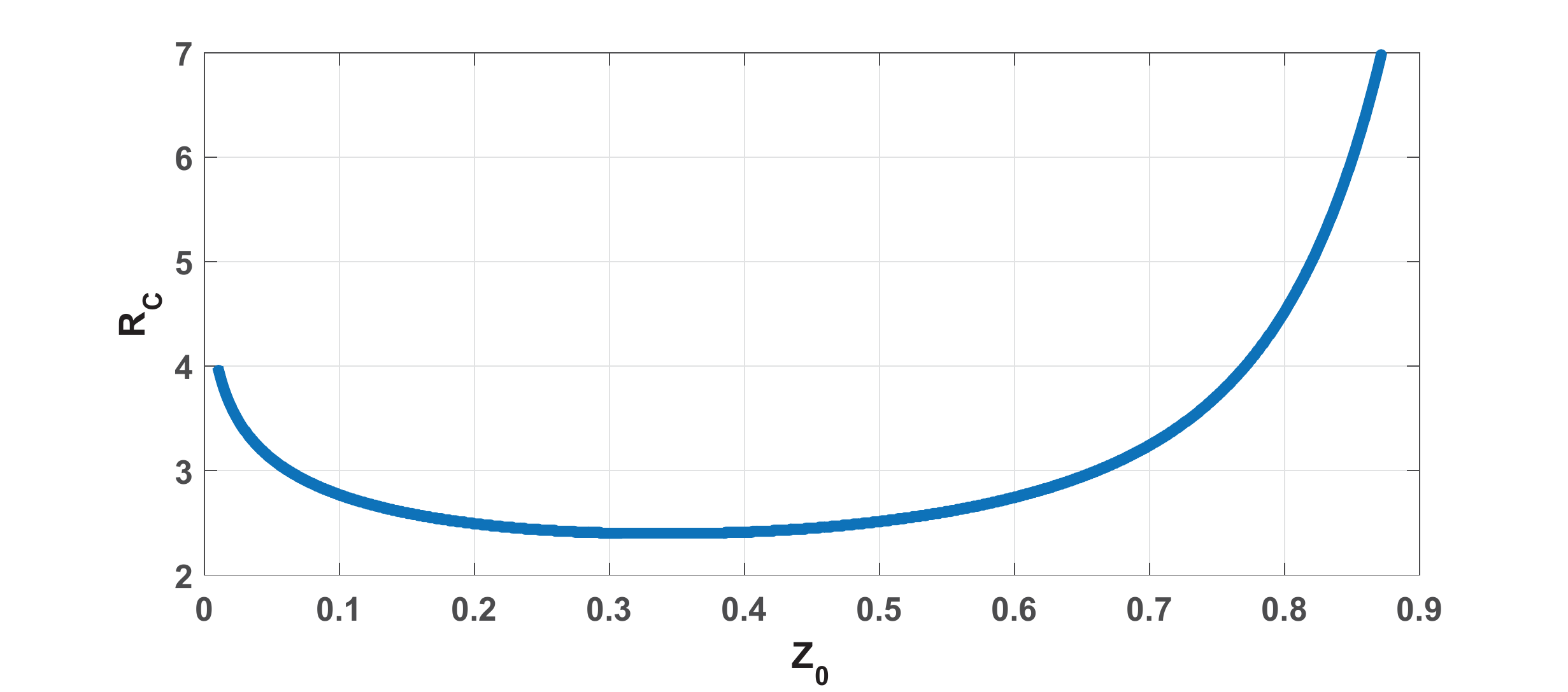}
		\caption{(Color online) (Top) Plot of the critical Rayleigh number as function of $Z_0$ for the case of an impervious lower boundary and a perfectly permeable upper boundary, where $\xi=1$ and $Da=0$, and the corresponding critical wavenumber. The minimum value of $R_C$ and $\alpha_C$ are $14.6646$ and $2.4$, respectively attained at $Z_0=0.38$. (Bottom) Plot of the critical wavenumber as function of $Z_0$; $\alpha_C \approx 5$ at $Z_0=0$ and increases with $Z_0$ to reach an asymptotic value $\approx 8$ as $Z_0 \rightarrow 1$} 
	\label{fig:2}
\end{figure}

\begin{figure}[h!]
 	\centering
 	\includegraphics[scale=.45]{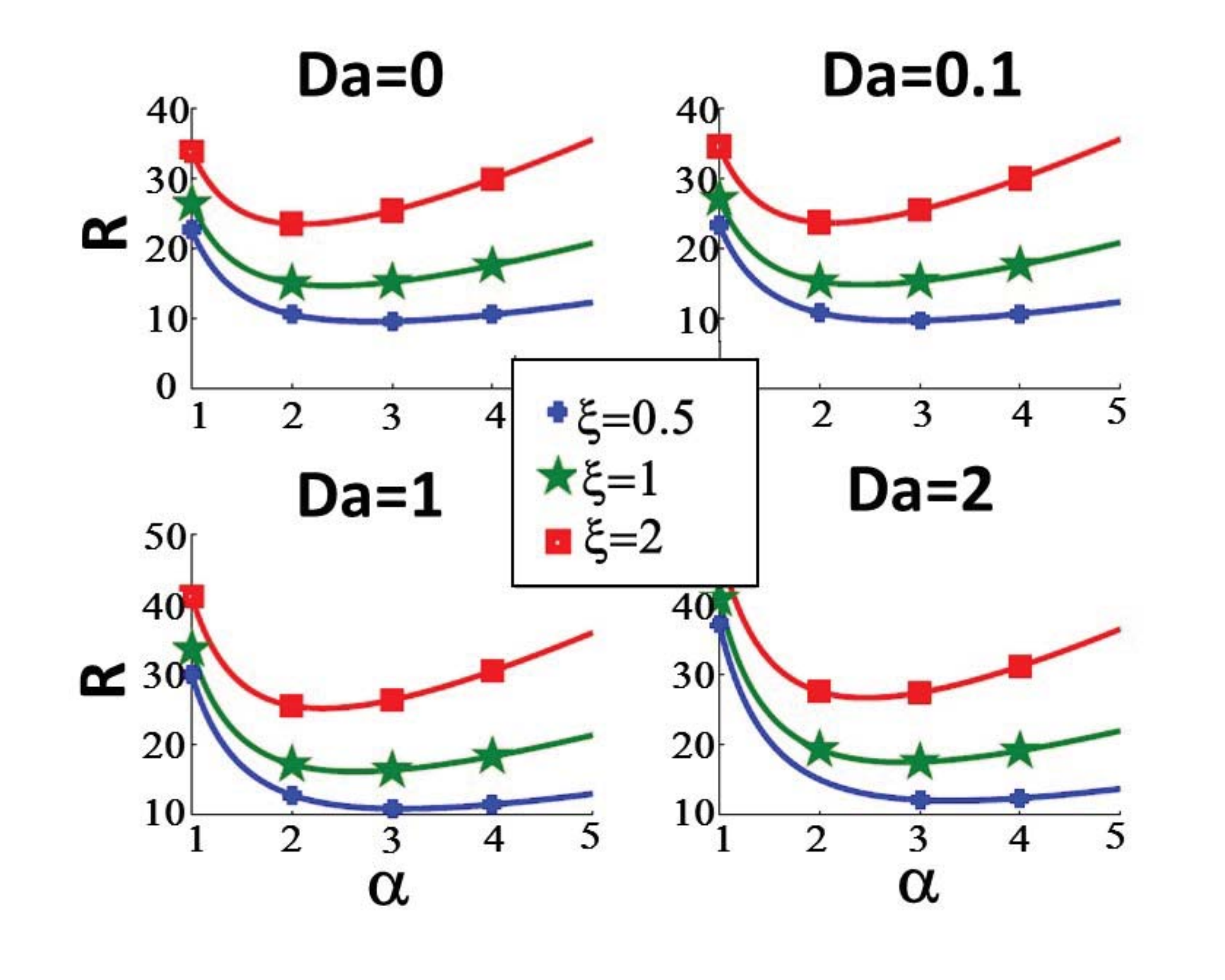}
 	\caption{Plot of the Rayleigh number $R$ versus $\alpha$ at $Z_0=0.38$ for the case of an impervious lower boundary and a perfectly permeable upper boundary for three distinct values of $\xi$, $\xi=0.5$, $\xi=1$, and $\xi=2$, and four different reaction rates, $Da=0$, $Da=0.1$, $Da=1.0$, and $Da=2.0$. The corresponding values of the thickness $Z_C$ are displayed in Table 1. }
	\label{fig:3} 
 \end{figure}
\begin{figure}[h!]
	\centering
	\includegraphics[scale=.4]{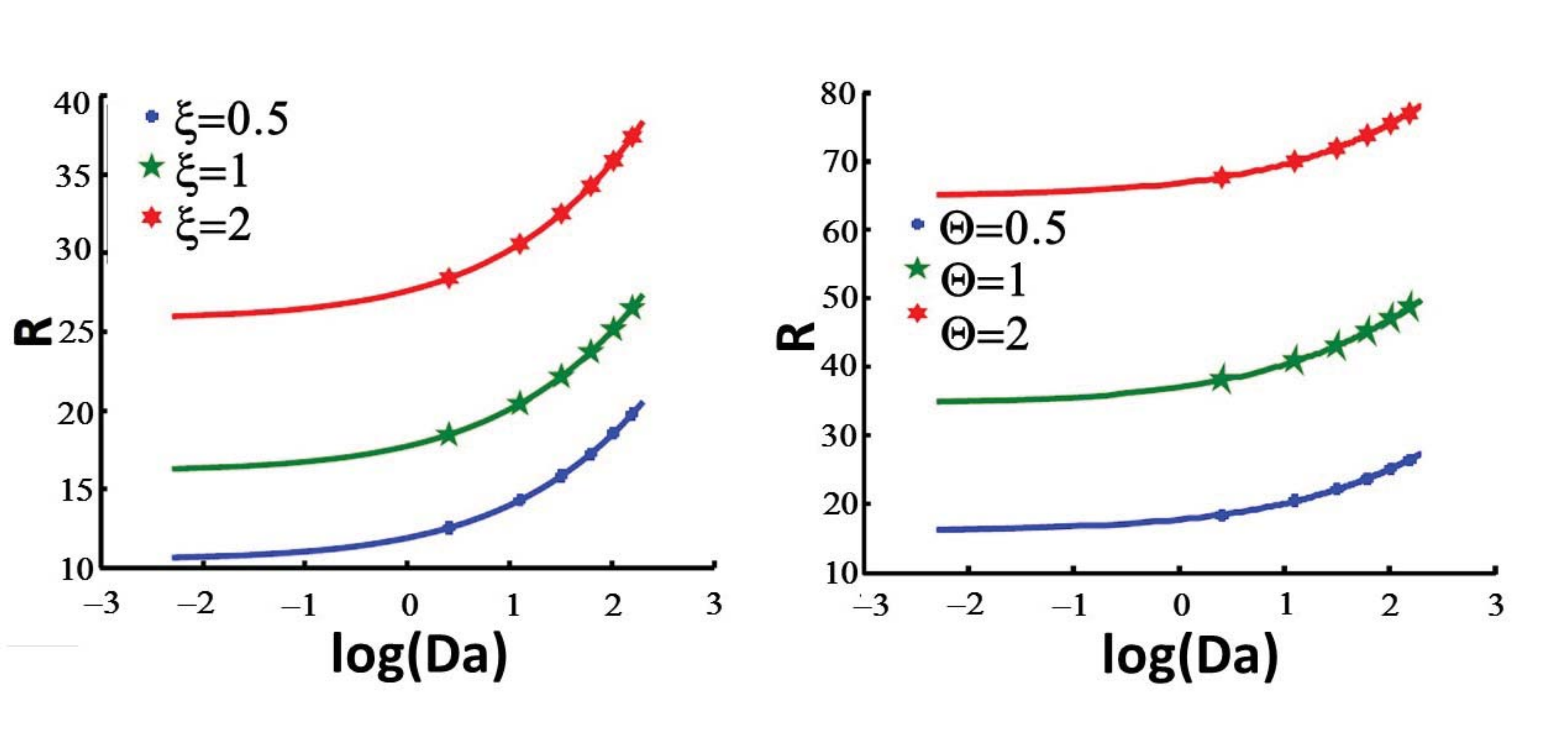}
	\caption{(left) The plot of critical Rayleigh number as a function of the log of the Damk\"{o}hler number with three distinct values of $\xi$, $\xi$=.5, $\xi$=1, and $\xi$=2. (right) $R$ versus $\log(Da)$ for three different permeabilities, $\Theta=0.1$, $\Theta=1.0$, and $\Theta=2.0$. The corresponding values of the thickness $Z_C$ are displayed in Tables $2$ and $3$.}
	\label{fig:9}
\end{figure}

Figure (\ref{fig:2}) depicts the dependence of the critical Rayleigh number and wavenumber on $Z_0$. For this set
of boundary conditions, namely a permeable top boundary and an impermeable lower
boundary, the minimum values are attained at $Z_0 = 0.38$. The system's stability characteristics,
the details of which are depicted in Figs. (\ref{fig:3},\ref{fig:6},\ref{fig:9}) and Tables (1-3), can
be summarized as follows: (i) the reaction acts to dissolve the $CO_2$ and thus has a
stabilizing effect, (ii) an increase in the disparity between the horizontal and vertical
diffusivities is associated with increased threshold values for instability onset to disturbances of shorter wavelength, (iii) the critical depth at instability onset, $Z_c$, increases
with $R_c$, and (iv) an increase in permeability acts to inhibit convection onset.
For example, we note from the plots of the Rayleigh number as function of the disturbance
wavenumber $\alpha$ shown in Fig. (\ref{fig:3}), that by varying $\xi$ and $Da$ we observe the
effects the diffusivity ratio and different reaction rates have on the system. As the vertical
diffusivity decreases, the $CO_2$ concentration is pushed closer to the top resulting
in a thinner layer of dense fluid becoming more stable.
 \\
 \\
 In order to determine concentration and velocity profiles of the system, we will solve the same homogeneous system, ${\bf Q}\,{\bf Y} =0$. However, we will normalize one of the unknowns. A new system of equations with $3$ unknowns is produced, 
 ${\bf {\hat Q}_1}\,{\bf Y1} ={\bf b}$. Plugging in the critical values previously found, the system is solved and the contours can be seen below for different reaction rates and diffusivity ratios.
 \begin{figure}[h!]
 	\centering
 	\includegraphics[scale=.4]{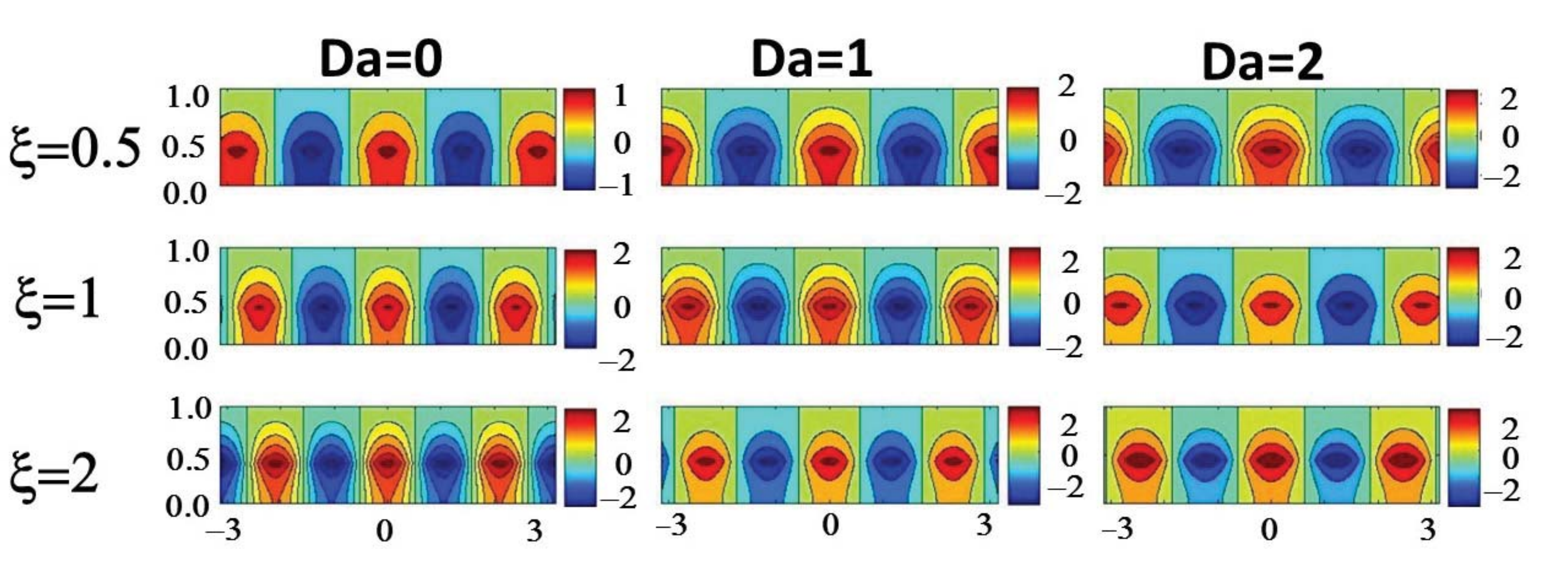}
 	\caption{Plot of the concentration perturbation contours at the onset of convection $(Z_c \approx 0.38)$ with three distinct values of $\xi$, $\xi$=0.5, $\xi$=1, and $\xi$=2, and three different reaction rates, $Da=0$, $Da=1$, and $Da=2$. The corresponding values of the thickness $Z_C$ are displayed in Table 1.}
 	\label{fig:4}
 \end{figure}
 \begin{figure}[h!]
 	\centering
 	\includegraphics[scale=.4]{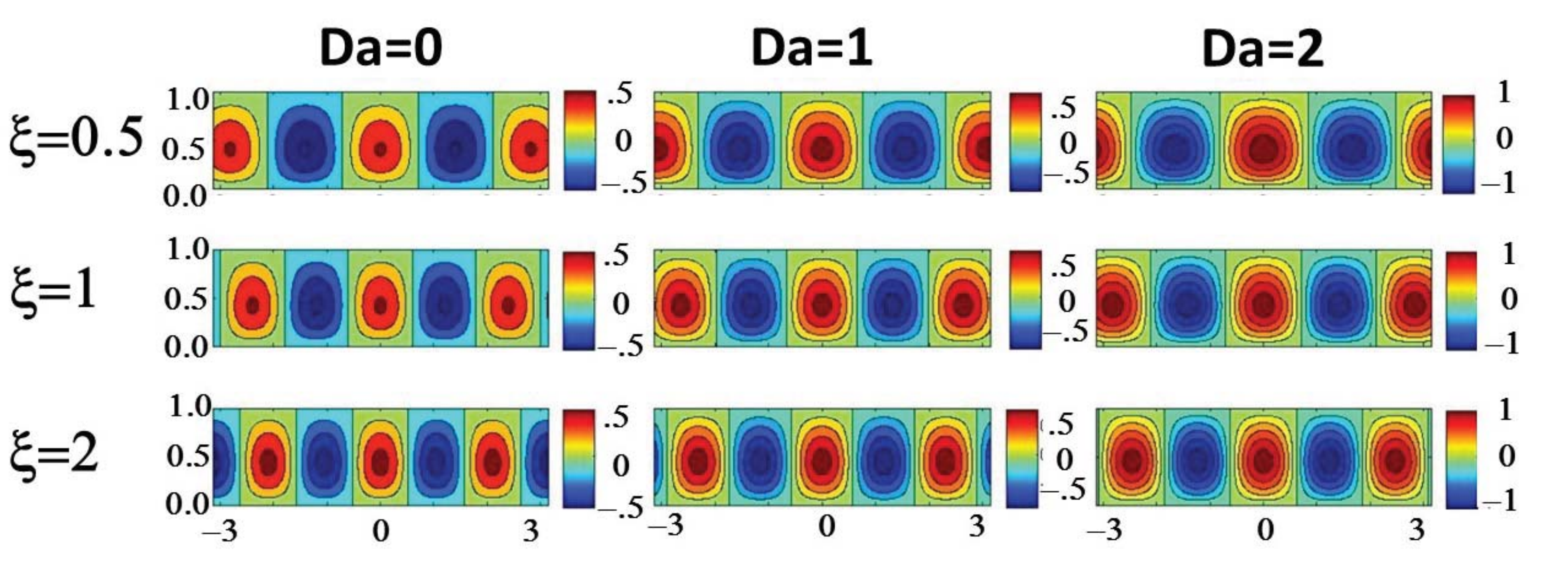}
 	\caption{Plot of $\phi$ at the onset of convection $(Z_C \approx 0.38)$ with three distinct values of $\xi$, $\xi=0.5$, $\xi=1$, and $\xi=2$, and three different reaction rates, $Da=0$, $Da=1.0$, and $Da=2.0$. The corresponding values of the thickness $Z_C$ are displayed in Table 1.}
 	\label{fig:5}
 \end{figure}
 The \enquote{fingering effect} from the Rayleigh-Taylor instability is seen in Fig. (\ref{fig:4}). One aspect of this plot to notice is that decreasing vertical diffusivity produces more \enquote{tongues}. This phenomenon can be explained by the resistance of vertical diffusion due to low vertical diffusivity in the darker regions i.e high concentration regions
 clump together forming these tongue structures. With regards to varying the vertical
 diffusivity in the velocity contours, as vertical $\xi$ increases the circulation currents
 spread out as expected. The increase in reaction rate produces more gradual concentration
 and velocity gradients as seen in Figs. (\ref{fig:4},\ref{fig:5}).\\

\begin{table}
\caption{Threshold values of the Rayleigh number, depth and wavenumber with varying reaction rates ($Da$) and diffusivity ratio ($\xi$).}
\begin{tabular}{|r|r|r|r|r|r|r|r|r|r|r|r|r|}
\hline
 & \multicolumn{4}{|c|}{$\xi=0.5$} & \multicolumn{4}{|c|}{$\xi=1$} & \multicolumn{4}{|c|}{$\xi=2$}\\
  \hline
$Da$  &  $0$   &  $0.1$   &   $1$  &  $2$  &  $0$   &  $0.1$   &   $1$  &  $2$   &  $0$   &  $0.1$   &   $1$  &  $2$\\
 \hline
$Z_c$ & $0.362$ & $0.363$ & $0.365$  &  $0.368$  &  $0.384$  &  $0.385$ & $0.394$ & $0.402$ & $0.41$ & $0.412$ & $0.43$ &  $0.443$ \\
\hline
$R_C$ & $9.56$ & $9.7$ & $10.87$ & $12.04$ &  $14.66$ & $14.83$ & $16.18$ & $17.5$ & $23.5$ & $23.6$ & $25.2$ & $26.7$ \\
\hline
$\alpha_C$ & $2.82$ & $2.85$ & $3.09$ & $3.32$ & $2.4$ & $2.43$ & $2.64$ & $2.83$ & $2.1$ & $2.12$ & $2.31$ & $2.47$ \\
\hline
\end{tabular}
\end{table}

\subsection{Depth-dependent permeability}
\begin{figure}[h!]
	\centering
	\includegraphics[scale=.5]{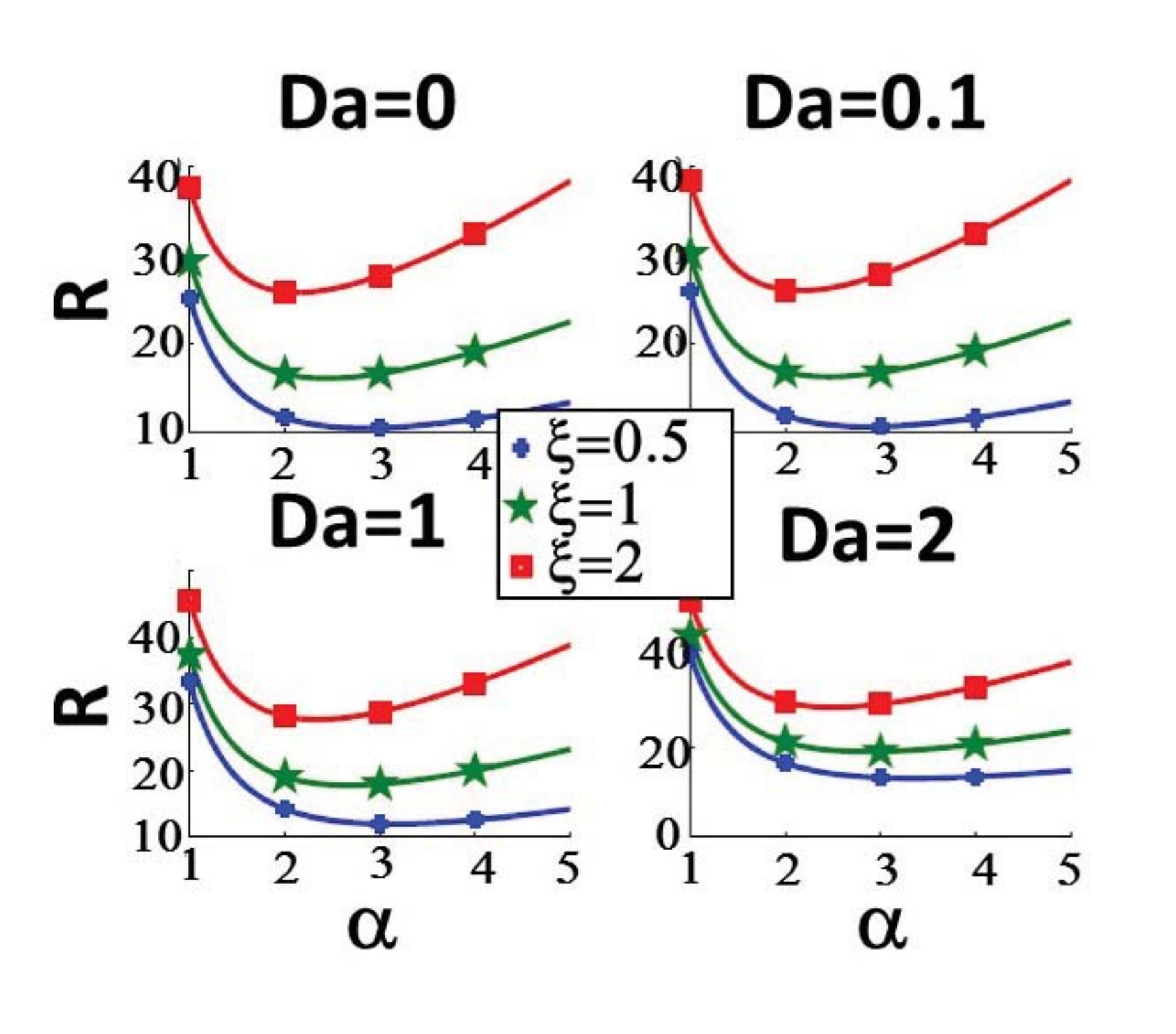}
	\caption{Plot of the Rayleigh number $R$ versus $\alpha$ for the case of an impervious lower boundary and a perfectly permeable upper boundary for three distinct values of $\xi$, $\xi=0.5$, $\xi=1$, and $\xi=2$, and four different reaction rates, $Da=0$, $Da=0.1$, $Da=1.0$, and $Da=2.0$ with $\Theta =0.1$. The corresponding values of the thickness $Z_C$ are displayed in Table $2$ and $3$. }
	\label{fig:6}
\end{figure}

In the previous section a simplification was made to the permeability. The model
function $f$ in Eq. (9) was assumed to be constant, when in reality it varies with respect
to depth. To capture this variation $f$ will be modeled as an exponential function in the vertical position within the aquifer
as $f(z) = \exp{-(1-\Theta\,z)}$. The coefficient $\Theta$, being specific to the rock formation of the sequestration site, is taken to vary between the values $0.1$ and $3$ in order to cover a wide range of idealized sites. The increase in
pressure will result in each subsequent layer to be more compressed. Pore space will
become more tightly packed resulting in decrease in permeability. An empirical exponential
curve fit was obtained by Klinkenberg \cite{25} that relates permeability variation
with pressure. Since pressure is depth dependent in this problem, an exponential variation
in permeability with depth can be obtained. To simplify our study, permeability
will only vary with $z$. According to Xu et. al \cite{7}, the impact of vertical permeability has a much stronger effect on the critical Rayleigh number, wavenumber, and critical
depth than the variation in horizontal permeability. After applying the exponential
substitution where $\Theta$ controls the variation in permeability, Eq. (9) reduces to
\\
\\
\begin{math}
	e^{-(1-\Theta z)} D^2W - \alpha^2\,e^{-(1-\Theta z)}\,W -D(e^{-(1-\Theta z)})\,DW = -e^{-2(1-\Theta z)}\,\alpha^2\,S
\end{math} \\
which upon simplification reduces to
\\
\begin{equation}\label{expdarcy}
(D^2-\Theta D-\alpha^2)W=-e^{-(1-\Theta z)}\alpha^2 S
\end{equation}
The transformed conservation of species remains the same
\begin{equation}
(D^2-p^2){S^+}=W\delta(z-Z_0)
\end{equation}
The solution method is the same as the previous section, once a solution to the ODE to Eq. (\ref{expdarcy}) is found.
The solution was initially found using $\Theta=0.1$. A plot of Rayleigh number versus wavenumber can be seen in Fig. (\ref{fig:6}).

\begin{figure}[h!]
	\centering
	\includegraphics[scale=.35]{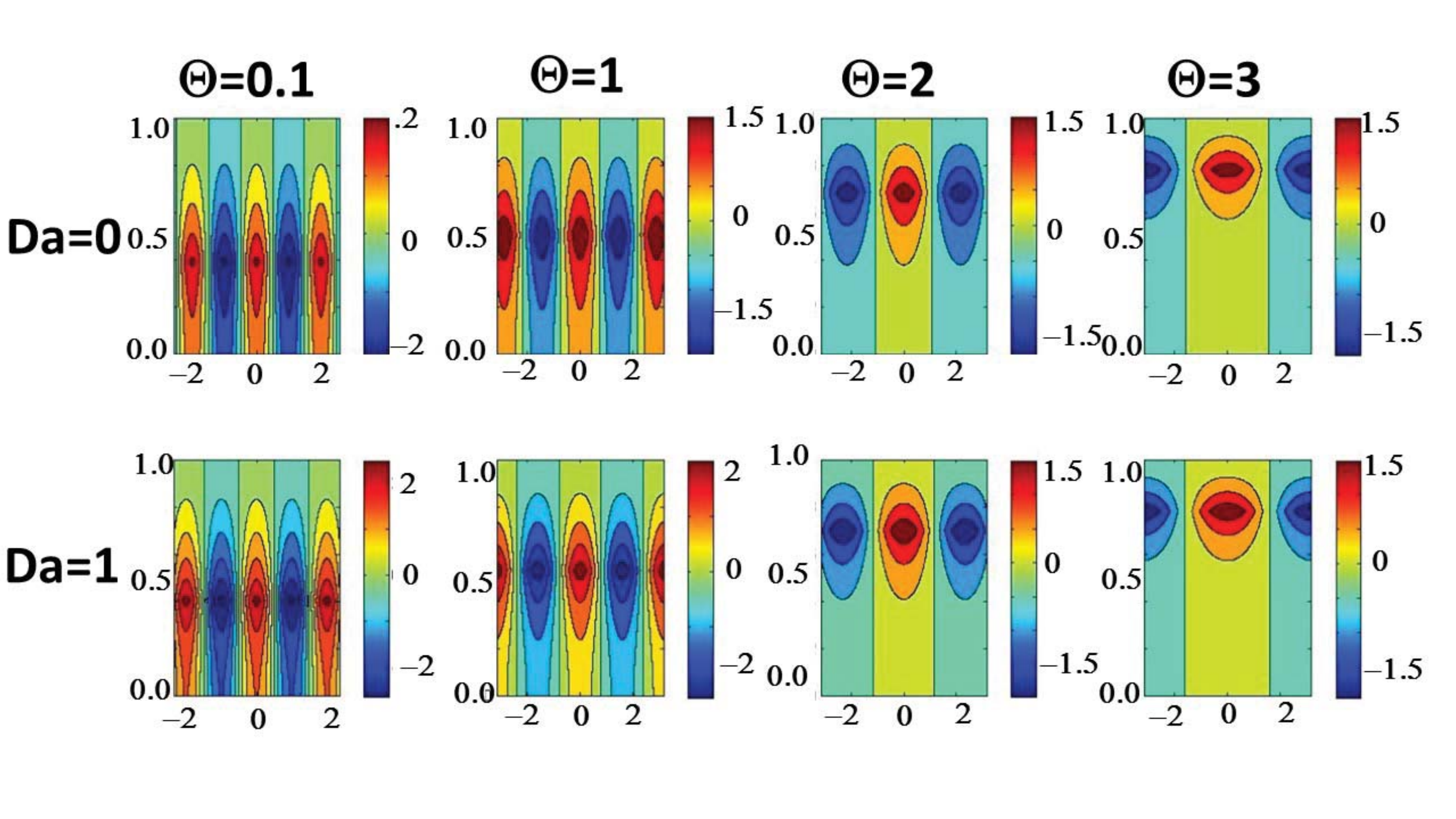}
	\caption{Plot of the perturbation contours at the onset of convection with four distinct values of $\Theta$, $\Theta$=.1, $\Theta$=1, $\Theta$=2, and $\Theta$=3, and two different reaction rates $Da=0$ and $Da=1.0$. The plots are all done at $\xi$=1.
	The corresponding values of the thickness $Z_C$ are displayed in Table $2$ and $3$.}
	\label{fig:7}
\end{figure}

\begin{figure}[h!]
	\centering
	\includegraphics[scale=.35]{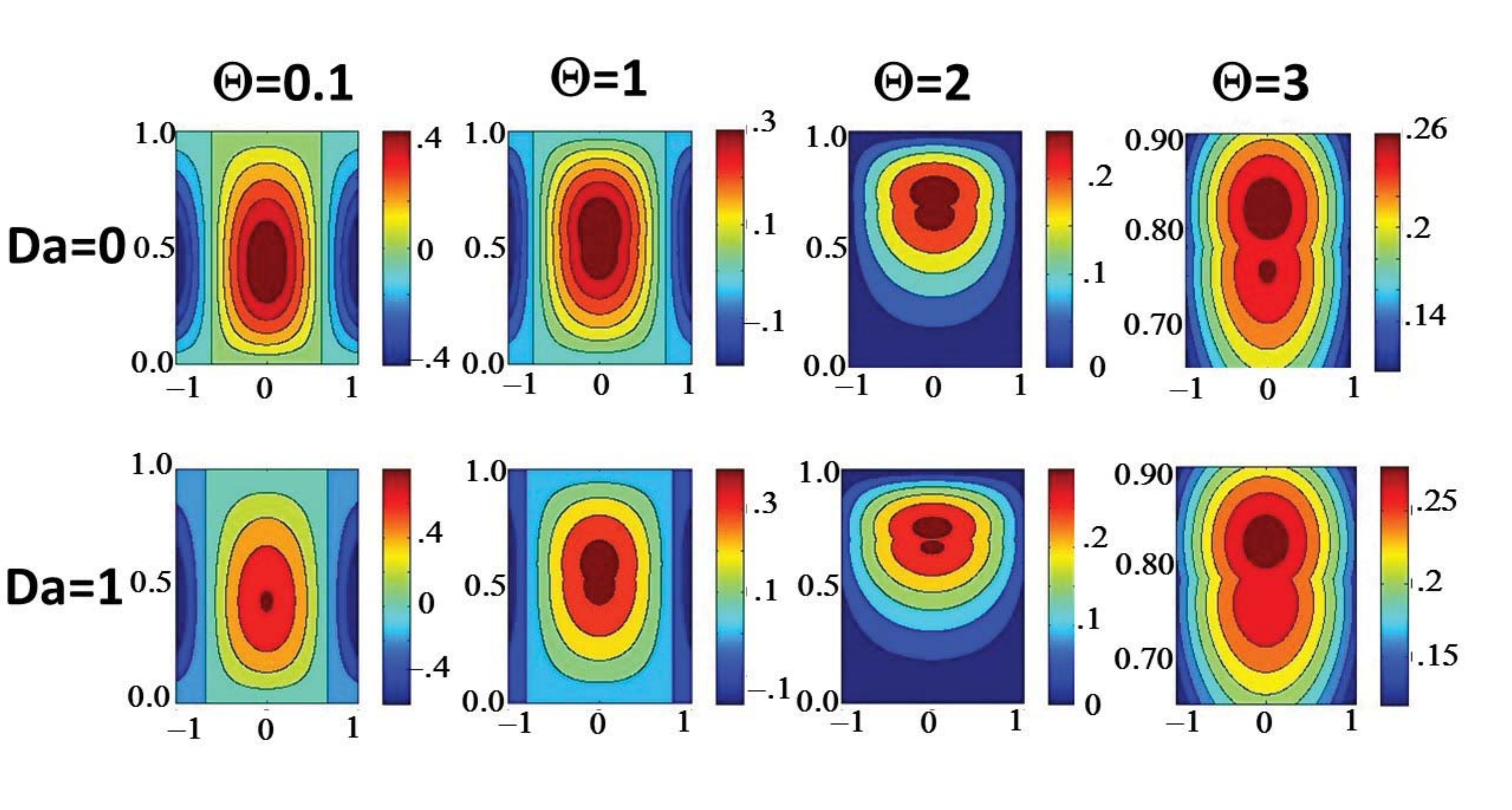}
	\caption{Plot of $\phi$ at the onset of convection with four distinct values of $\Theta$, $\Theta$=.1, $\Theta$=1, $\Theta$=2, and $\Theta$=3, and two different reaction rates $Da=0$ and $Da=1.0$. The plots are all done at $\xi$=1. The corresponding values of the thickness $Z_C$ are displayed in Table 2.}
	\label{fig:8}
\end{figure}

\begin{table}
	\caption{Threshold values of the Rayleigh number, depth and wavenumber with varying reaction rates ($Da$) and diffusivity ratio ($\xi$) for $\Theta$=0.1.}
	\begin{tabular}{|r|r|r|r|r|r|r|r|r|r|r|r|r|}
		\hline
		& \multicolumn{4}{|c|}{$\xi=0.5$} & \multicolumn{4}{|c|}{$\xi=1$} & \multicolumn{4}{|c|}{$\xi=2$}\\
		\hline
		$Da$  &  $0$   &  $0.1$   &   $1$  &  $2$  &  $0$   &  $0.1$   &   $1$  &  $2$   &  $0$   &  $0.1$   &   $1$  &  $2$\\
		\hline
		$Z_c$ & $0.369$ & $0.37$ & $0.376$  &  $0.383$  &  $0.392$  &  $0.393$ & $0.403$ & $0.412$ & $0.419$ & $0.42$ & $0.432$ &  $0.441$ \\
		\hline
		$R_C$ & $10.49$ & $10.64$ & $11.9$ & $13.15$ &  $16.1$ & $16.28$ & $17.73$ & $19.15$ & $25.77$ & $25.97$ & $27.59$ & $29.17$ \\
		\hline
		$\alpha_C$ & $2.84$ & $2.87$ & $3.12$ & $3.35$ & $2.44$ & $2.46$ & $2.67$ & $2.87$ & $2.14$ & $2.16$ & $2.34$ & $2.5$ \\
		\hline
	\end{tabular}
\end{table}
A very similar trend to Fig. (\ref{fig:3}) is shown. A table is needed to see that exact numerical changes in the critical values since Fig. (\ref{fig:6}) is so close to Fig. (\ref{fig:3}).
Since the values of different critical numbers for $\Theta=0.1$ and $\Theta=0$ are visually negligible, concentration and velocity contours will be shown for different increasing $\Theta$. 
For large $\Theta$ values, we observe that the concentration gradients begin to steepen at
the critical depth as shown in Fig. (\ref{fig:7}). This is expected since permeability becomes
exponentially large, resulting in a zero mass flux past a certain depth. As the permeability
increases the finger-like structures disappear and evolve to pure tongue-like
structures. Once permeability reaches a certain value, the $CO_2$ mass transfer can no
longer occur, thus completely obstructing the flow. Once the flow is completely obstructed
there will no longer be a concentration gradient or velocity gradient; which
is seen in Figs. (\ref{fig:7},\ref{fig:8}) when $\Theta=2$ and $\Theta=3$. Increasing of permeability shifts the $CO_2$
concentration toward the top as expected. In both contour plots, the size of each
concentration cell and convection cell are reduced once reaction rates are present.
Physically, reaction rates represent the dissolution of $CO_2$ into the surrounding brine.
The result from $CO_2$ dissolution taking place is an overall reduction in the overturning
motion due to the buoyancy-driven instabilities . This is indicative of a more
uniform distribution of $CO_2$ concentrations and velocity profiles in the domain. Reaction
rates act as a dampener for both concentration and velocity gradients seen in
Figs. (\ref{fig:7},\ref{fig:8}). Figure (\ref{fig:8}) depicts the formation of double convection cells. As variation
in permeability increases the convection cells compact around $Z_0$ until they split
into two separate cells. As variation in permeability increases the convection cells
become squeezed around $Z_0$ until they split into two separate cells. The high permeability
effects push the velocity cells toward the top of the domain, until one of the
cells is completely above the critical depth. Once this happens, the convective motion
induces a smaller convection cell below the critical depth. Since the signs of the
vorticity in each cell are the same, the convective motions are solutally coupled (due
to variations in density). Permeability increase separates the largest convection cell
into double layer convection cells, similar to the study done by Rasenat et. al \cite{26}.
Their study investigates the cyclic variation between thermal and viscous coupling
in double layer convection. However in our study, permeability and density that vary
with depth do not seem to induce any penetrative convection due to viscous viscous
coupling. For the case of $\Theta=2$ in Fig. (\ref{fig:8}), the dampening effect of chemical reactions
taking place causes the velocity cell to shrink. The cell becomes small enough to concentrate
above the critical depth but still large enough to induce a smaller cell below
the critical depth. For $\Theta=3$ the cell is not large enough to induce another convection
cell below the critical depth.\\
\\    

\begin{table}
	\caption{Threshold values of the Rayleigh number, depth and wavenumber with varying reaction rates ($Da$), varying permeability ($\Theta$), and $\xi$=1.}
	\begin{tabular}{|r|r|r|r|r|r|r|r|r|r|r|r|r|}
		\hline
		& \multicolumn{5}{|c|}{$Da=0$} & \multicolumn{5}{|c|}{$Da=1$}\\
		\hline
		$\Theta$  &  $.01$   &  $0.1$   &   $1$  &  $2$  &  $3$   &  $0.01$   &   $.1$  &  $1$   &  $2$   &  $3$ \\
		\hline
		$Z_c$ & $0.385$ & $0.392$ & $0.504$  &  $0.684$  &  $0.78$  &  $0.394$ & $0.403$ & $0.531$ & $0.699$ & $0.781$ \\
		\hline
		$R_C$ & $14.80$ & $16.10$ & $34.60$ & $64.87$ &  $97.35$ & $16.33$ & $17.73$ & $37.00$ & $66.75$ & $98.73$ \\
		\hline
		$\alpha_C$ & $2.41$ & $2.44$ & $2.87$ & $4.3$ & $6.25$ & $2.64$ & $2.67$ & $3.18$ & $4.64$ & $6.39$ \\
		\hline
	\end{tabular}
\end{table}

\subsection{Linear Analysis with nearly-impermeable $CO_2$ flux upper boundary}
This section is devoted to the linear stability analysis of the second boundary condition case, the results of which will used in the weakly non-linear analysis. The analysis is limited to the case of constant permeability and isotropic diffusion. Equations (8, 9) can be combined into a single equation for the velocity, namely
\begin{equation}
\label{stab1}
(D^2 - \alpha^2)^2W = \alpha^2\,R\,\delta{(z-Z_0)}\,W,  
\end{equation}
with boundary conditions that correspond to rigid and nearly-impermeable to solute flow. Due to the presence of the delta function term,
we expect the functions $W(z)$ and $S(z)$ to be $C^2([0,1])$ and $C([0,1])$, respectively. We solve Eq.(\ref{stab1}) in the two separate regions, $0<z<Z_0$ and $Z_0<z<1$ which we label with the super scripted variables $W^-(z)$ and $W^+(z)$, respectively. We have
\begin{equation}
\label{velocity1}
(D^2 - \alpha^2)^2 W^-=0, \quad W^-(0)=0, \quad DW^-(0) -\alpha^2\,DW^-(0)=0
\end{equation}
the solution of which is given by
\begin{equation}
W^-(z)= (A^- + B^-\,z)\sinh{(\alpha z)}
\end{equation}
where $A^-$ and $B^-$ are constants to be determined later by matching at $z=Z_0$ with the corresponding solution in the upper layer. The latter is found by solving Eq. (\ref{velocity1}) with the boundary conditions $W^+(1)=0$ and  $D^3W^+(1) - \beta\,D^2W^+(1)-\alpha^2\,DW^+(1)=0$. It is given by
\begin{equation}
W^+(z)=(\beta/\alpha)B^+\,(z-1)\,\cosh{\alpha(z-1)}+(A^++B^+\,(z-1))\,\sinh{(\alpha(z-1))}
\end{equation}
 solved using the same techniques used in the previous section. A solution for $W^-$ and $W^+$ is found to be
\begin{equation}
W^-(z)= (A^- + B^-\,z)\sinh{(\alpha z)}
\end{equation}
\begin{equation}
W^+(z)=(\beta/\alpha)B^+\,(z-1)\,\cosh{\alpha(z-1)}+(A^++B^+\,(z-1))\,\sinh{(\alpha(z-1))}
\end{equation} 
\begin{figure}[h]
	\centering
	\includegraphics[scale=0.3]{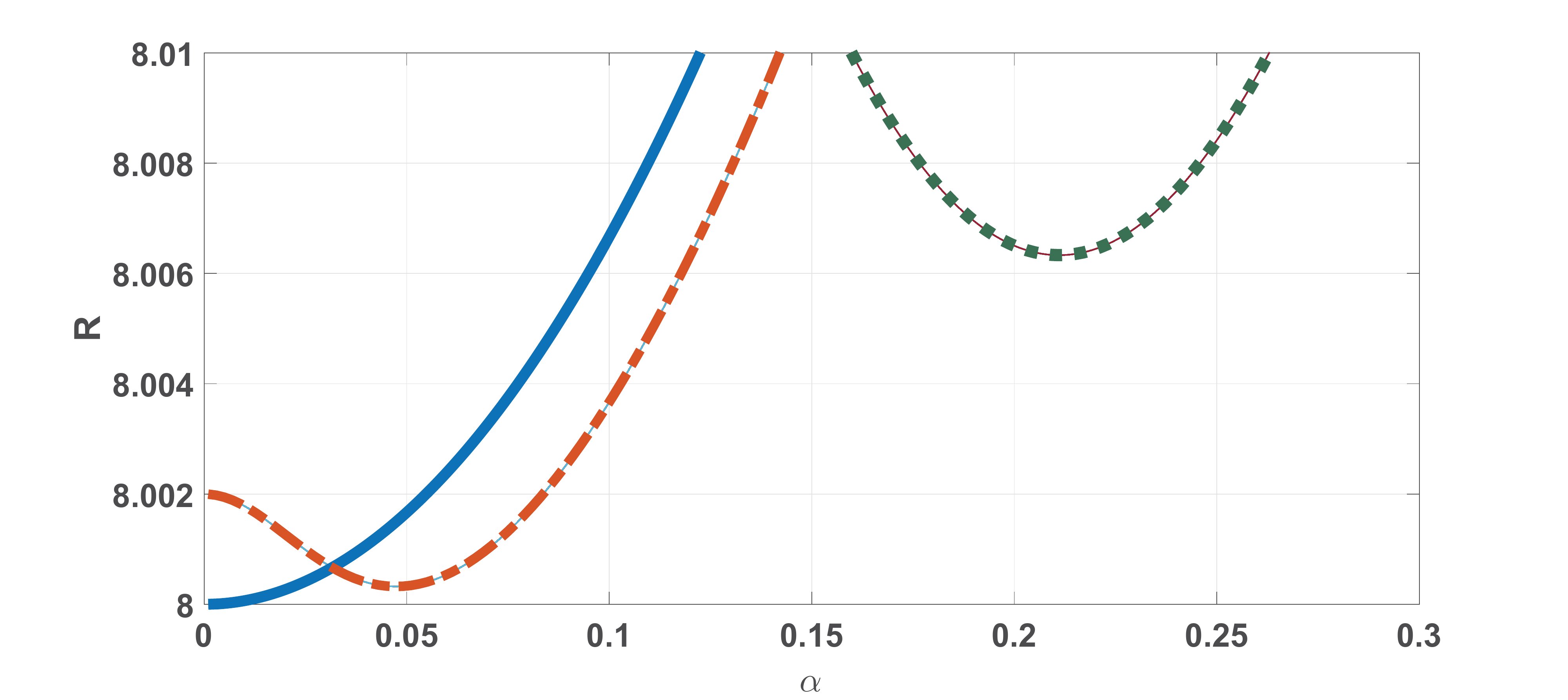}\\
	\caption{(Color online)  Plot of the Rayleigh number for disturbances of wavenumber $\alpha$ for $Z_0=0.5$ and a set of three values of the dimensionless mass transfer coefficient, $\beta=0$ (continuous line), $\beta=0.01$ (dash-dotted line) and $\beta=0.02$
	 (dotted line)} 
	\label{fig:10}
\end{figure} 
 We observe that the minimum Rayleigh number $R_C$ approaches the value $8$ and the corresponding wavenumber $\alpha_C$ approaches $0$ as $\beta \to 0$. For the same boundary conditions and a fluid layer that is continuously stratified, $R_C=12$ and 
 $\alpha_C=0$ \cite{13}. We see that the step-model function approach gives a more conservative value for the critical numbers.

 \section{Weakly nonlinear analysis}
 
\begin{figure}[h]
	\centering
	\includegraphics[scale=0.35]{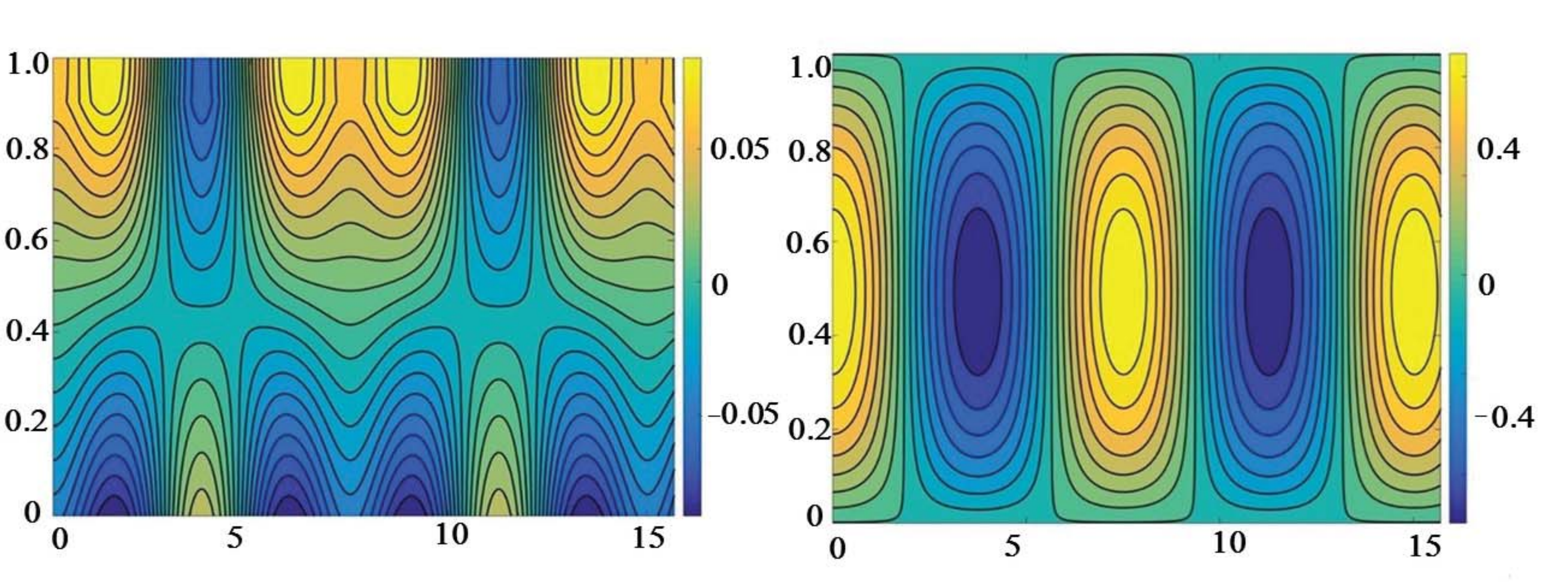}
	\caption{(Left) Plot of the concentration perturbation contours, $c_2$. (Right) Plot of the streamlines, $\phi_0$, at onset for ${\hat \mu}=0.65$, $\Gamma =1.0$, $\delta=0.1$ and, ${\hat \beta} = $12 for an interface at $Z_0=0.5$.}
	\label{fig:11}
\end{figure}

\begin{figure}[h]
	\centering
	\includegraphics[scale=0.35]{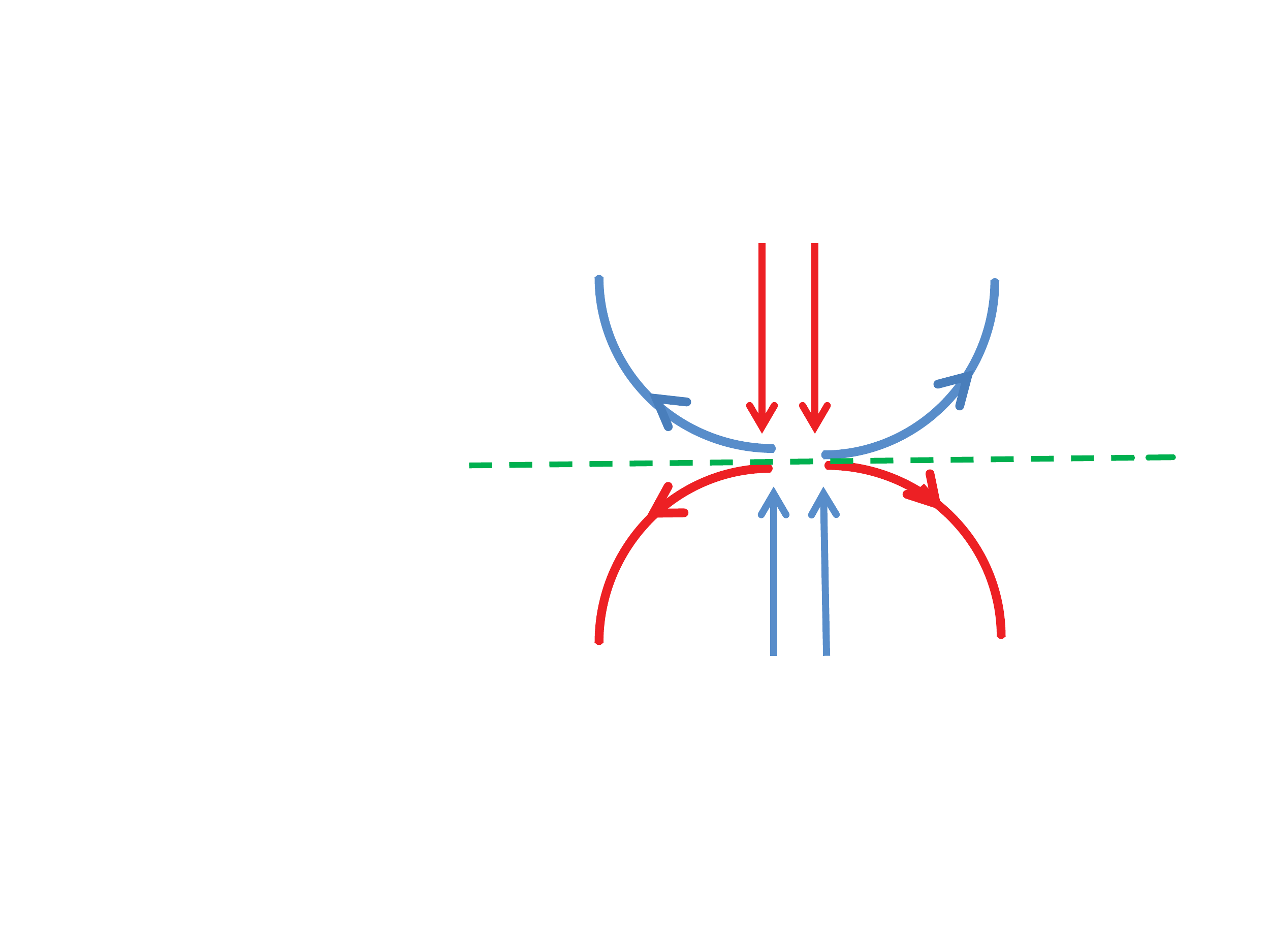}
	\caption{A schematic diagram of the order $\delta^2$ concentration where a carbon-free plume ascend and splits into two at the interface to leave room for the carbon-rich plume to descend as it also splits into two at the interface.}
	\label{fig:12}
\end{figure}

 We derive an evolution equation for the leading order concentration profile by making use of long wavelength asymptotics valid in the limit $\beta$ approaches zero, and as depicted in Fig. (\ref{fig:10}), the critical wavenumber approaches zero as $\beta$ approaches zero.
 We invoke the analysis in \cite{27} which has shown that the scaling of the critical wavenumber with the mass transfer coefficient 
 $\beta$ takes the form, $\alpha_C^4 \sim (\beta/h)$ as $\beta \to 0$, for a dimensionless thickness of the plates of order unity. Thus, the instability onset
 is characterized by the scaling $\alpha_C^2 \sim \sqrt{\beta}$ as ${\beta \to 0}$. We introduce the small parameter $\epsilon$ and let $\beta = \epsilon^4\,{\hat r}$ where ${\hat r}$ is $O(1)$. Near the onset of convection, the characteristic dimension of the convection cell is expected to be much larger than the fluid layer height. Thus, we scale the horizontal gradient by
 $\epsilon$, $0 < \epsilon \ll1$,  $\partial /\partial x= \epsilon\,\partial/\partial X$ and maintain the vertical direction unscaled. The slow time is scaled as $\tau = \epsilon^4\,t$ and we express the supercritical Rayleigh number as $R = R_c + \mu^2\,\epsilon^2$ where $\mu = O(1)$ \cite{28}. After, for convenience in the calculations, we replace $c$ by $-c$, the system of equations Eqs. (\ref{system2},\ref{system3}) reduces to
 \begin{eqnarray}
 \label{NL1}
 D^2\phi + \epsilon^2\,\phi_{xx} &= &- (R_c + \epsilon^2\,\mu)c\\
 \label{NL2}
 D^2 c + \epsilon^2 c_{XX} + \epsilon^2\,\phi_{XX}\,\delta{(z-Z_0)} & = & \epsilon^4\,c_{\tau} + \epsilon^2\,(D\phi)_X\,c_X - \epsilon^2\,\phi_{XX}\,Dc
 \end{eqnarray}  
 We expand the variables $c$ and $\phi$ in a power series of $\epsilon^2$ as follows,
 \begin{equation}
 \label{NL3}
 c=\sum_{n=0}^{\infty} \epsilon^{2n}\,c_{2n}, \qquad \phi=  \sum_{n=0}^{\infty} \epsilon^{2n}\,\phi_{2n}
 \end{equation}
 which are then substituted into Eqs. (\ref{NL1},\ref{NL2}) to yield a series of boundary values problems of different orders in $\epsilon$ that are solved in a successive manner. At the leading order, we have
 \begin{equation}
 D^2 c_0=0, \qquad Dc_{0}(0)=Dc_{0}(1)=0,
 \end{equation}
 whose solution is $c_0 =h(X,\tau)$, and
 \begin{equation}
 D^2\phi_{0} = -R_C\,h, \qquad \phi_0(0)=\phi_0(1)=0,
 \end{equation}
 whose solution is given by $\phi_{0}(X,z,\tau)=-R_C\,h\,P(z)$ where $P(z) = (z^2-z)/2$. Proceeding to the next order in $\epsilon$, 
 Eq.(\ref{NL2}) yields
 \begin{equation}
 \label{NL4}
 D^2 c_2 + h_{XX} = -R_C\,(h_X)^2\,DP + (\phi_0)_{XX} \delta{(z-Z_0)}
 \end{equation}
 Due to the presence of the delta function term in Eq. (\ref{NL4}), the problem is solved for ${c_2}^-$ in the region $0 \le z < Z_0$ and for ${c_2}^+$ in $Z_0 < z \le 1$ as follows,
 \begin{eqnarray}
 D^2 {c_2}^- &=& -h_{XX} - R_C (h_X)^2\,DP, \qquad D{c_2}^-(0)=0,\cr
 \label{NL5}
 D^2 {c_2}^+ &=& -h_{XX} - R_C (h_X)^2\,DP, \qquad D{c_2}^+(1)=0,
 \label{NL6}
 \end{eqnarray}
 We have
 \begin{eqnarray}
 {c_2}^-& = &-(z^2/2) h_{XX} - R_C(h_X)^2\,P_1(z) + {\cal G}(X)\cr
 \label{NL7}
 {c_2}^+& = & -(z^2/2 - z) h_{XX} - R_C (h_X)^2\,P_1(z) + {\cal E}(X)
 \label{NL8}
 \end{eqnarray}
 where $P_1(z)$ is an anti-derivative of $P(z)$ and ${\cal A}(X)$ and ${\cal B}(X)$ are arbitrary functions of $X$ which are related 
 by ${\cal G} = {\cal E} + Z_0\,h_{XX}$ obtained by imposing the continuity of $c$ at $Z_0$. Thus, Eqs. (\ref{NL7}) reduce to
\begin{eqnarray}
{c_2}^- &= &-(z^2/2-Z_0)h_{XX} - R_C (h_X)^2\,P_1(z) + {\cal E} \cr
{c_2}^+ & = & -(z^2/2 -z) h_{XX} - R_C (h_X)^2\,P_1(z) + {\cal E}
\end{eqnarray}
Finally, we impose an orthogonality condition, $< c_0\,c_2> =0$ where $< \bullet>$ stands for integration with respect to $z$ from $z=0$ to $z=1$. We find ${\cal E} = -R_C (h_X)^2/24 - (3 Z_0^2 +2) h_{XX}/6$. Hence
\begin{eqnarray}
{c_2}^- &=& -[(3z^2+3 {Z_0}^2-6Z_0+2)/6] h_{XX} - R_C\,[(4z^3-6 z^2+1)/24](h_X)^2 \cr
\label{NL9}
{c_2}^+ &=& -[3z^2 + 3 {Z_0}^2 - 6z +2)/6] h_{XX} - R_C\,[(4z^3-6z^2+1)/24](h_X)^2
\label{NL10}
\end{eqnarray}
Upon integrating Eq.(\ref{NL4}) from $Z_0-\ell$ to $Z_0+\ell$ and using Eqs. (\ref{NL9}) and (\ref{NL10}), we find upon taking the limit $\ell$ approaches zero that the critical Rayleigh number is given by
\begin{equation}
\label{RC}
R_C = {2 \over Z_0 (1-Z_0)}
\end{equation}
The value of $R_C$ and its dependence on $Z_0$ are in agreement with the numerical results, Fig. (\ref{fig:3}) and it attains the minimum value of $R_C=8$ for $Z_0=0.5$.   Equating terms of order $\epsilon^2$ in Eq. (\ref{NL1}) leads to the boundary value problem
\begin{equation}
\label{NL11}
D^2 \phi_2 = R_C P(z) h_{XX} -\mu^2\,f - R_C\,c_2, \qquad \phi_2(0)=\phi_2(1)=0
\end{equation}
which we again solve in the two regions corresponding to ${c_2}^-$ and ${c_2}^+$. We find
\begin{eqnarray}
{\phi_2}^- & = & {S_1}^-(z)\,h_{XX} + {S_2}^{-}(z) (h_X)^2 - (\mu^2/2)z^2\,h + {\cal C}^-\,z, \cr
\label{NL12}
{\phi_2}^+ & = & {S_1}^+(z)\,h_{XX} + {S_2}^{+}(z) (h_X)^2 - (\mu^2/2)(z^2-1)\,h + {\cal C}^+\,(z-1)
\label{NL13}
\end{eqnarray}
where
\begin{eqnarray}
\label{NL14}
{S_1}^-(z) &=& R_C [ 2z^6 + 5(6{Z_0}^2 -12 Z_0 +16) z^4 - 120 z^3]/1440, \cr
{S_2}^-(z) &=& {R_C}^2 (4z^7 -14z^6 +35 z^4)/20160, \cr
{S_1}^+(z) &=& R_C [z^6 - 6z^5 +(15 {Z_0}^2 + 40) z^4 - 60 z^3 - 15 {Z_0}^2 + 25]/720, \cr
{S_2}^+(z) &=& (4 z^7 - 14z^6 + 35 z^4 -25)/20160
\end{eqnarray}
where ${\cal C}^-$ and ${\cal C}^+$ are determined by imposing the continuity of ${\phi_2}$ and $D\phi_2$ at $Z_0$. We find,
\begin{eqnarray}
\label{DC2}
{\cal C}^- &=& R_C [ (-66 {Z_0}^5 + 90 {Z_0}^4 - 15 {Z_0}^2 + 25)/720] h_{XX} -(5 {R_C}^2/4032) (h_X)^2 +(\mu^2/2)h \cr
{\cal C}^+&=& R_C [ (- 66 {Z_0}^5 - 15 {Z_0}^2+25)/720] h_{XX} -(5 {R_C}^2/4032) (h_X)^2 + (\mu^2/2)h
\end{eqnarray}
Finally, the order $\epsilon^4$ of Eq. (\ref{NL2}) yields
\begin{equation}
\label{Evo1}
D^2 c_4 = -(c_2)_{XX} + c_{\tau} + (D \phi_2)_X\,h_X - (\phi_0)_{XX} Dc_2 + (\phi_2)_{XX}\,\delta{(z-Z_0)}.
\end{equation}
We do not solve for $c_4$ but rather integrate Eq. (\ref{Evo1}) over the interval $(0,1)$ to obtain an equation for $h$ that describes the space and time evolution of $h$ near the onset of the instability. Upon setting $Z_0=1/2$ and $R_C=8$, we find, 
\begin{equation}
\label{Evo}
h_{\tau} = (-43/576)\,h_{XXXX}-(\mu^2/8)\,h_{XX} - {\hat r}\,h + (29/120)\,(h_X^2)_{XX} + (8/5)\,(h_X^2)\,h_{XX}.
\end{equation}
We introduce the following scales and variable transformations, $h=a \eta$, $\xi = b\,X$ and ${\hat \tau} = e \tau$,
where
\begin{align}
& a=\sqrt{215/4608}, \quad b=[(576/43)\sqrt{4608/215}]^{1/4}, \\
&\quad {\hat \beta} = a\,{\hat r}, \quad {\hat \mu} = \mu^2\,a b^2/16, \quad
e=1/a
\end{align}

to express Eq. (\ref{Evo}) in the following reduced form,
\begin{equation}\label{Reduced}
{\partial \eta \over \partial {\hat \tau}} = - \eta_{\xi\xi\xi\xi} -2 {\hat \mu}^2\,\eta_{\xi\xi} - {\hat \beta}\eta
+\eta_{\xi\xi}(\eta_\xi)^2 + \Gamma (\eta_\xi \eta_{\xi\xi\xi} + (\eta_{\xi\xi})^2)
\end{equation}
were $\Gamma = 1309/468$. Except for the value of the coefficient $\Gamma$, this is the same equation that is typically found using long wavelength asymptotics in convection between poorly conducting boundaries. A detailed derivation and numerical analysis of the equation  are described in \cite{28}. The trivial solution to Eq. (\ref{Reduced}) pertains to the static solution, the stability of which is investigated by considering its linear part. Upon introducing the normal modes
$ \eta(\xi,{\hat \tau}) = \exp{(\sigma\,{\hat \tau} + \imath\,\gamma\,\xi})$, we obtain the dispersion relation
\begin{equation}
\label{Disp2}
\sigma = -(\gamma^2 - {\hat \mu}^2)^2 + {\hat \mu}^4 - {\hat \beta}
\end{equation}
Thus, the static state, $\eta=0$, is unstable whenever $ {\hat \beta} < {\hat \mu}^4 $. We investigate the weakly nonlinear evolution of this instability by conducting a perturbation analysis around the linear solution. We quantify the deviation from the linear instability threshold by  the small parameter $\delta$, the sign of which can be either positive or negative to account for the possible occurrence of either sub-critical or super-critical instability. Thus, for $ |\delta| \ll 1$, we expand ${\hat \beta}$ and $\eta$ as follows
\begin{equation}
{\hat \beta} = {\hat \mu}^4-\delta \beta_1 - \delta^2\,\beta_2, \quad
\eta = \delta \eta^{(0)} + \delta^2\,\eta^{(1)} + \delta^3 \eta^{(2)} \ldots
\end{equation}
in Eq. (\ref{Reduced}) and seek its periodic solutions. At order $\delta$, we have
\begin{equation}
\label{ord1}
\eta_{\xi\xi\xi\xi}^{(0)} + 2 {\hat \mu}^2 {\eta_{\xi\xi}}^{(0)} + {\hat \mu}^4 \eta^{(0)}=0
\end{equation}
whose normalized solution on the interval $(-\pi/{\hat \mu}, \pi/{\hat \mu})$ is $\eta^{(0)} = \sqrt{2}\,\cos{({\hat \mu}\xi)}$.
We expect that mixed secular terms will appear at $O(\delta)$ and $O(\delta^2)$ due to the linear part ${\cal L}$ and the nonlinear term 
$\eta_{\zeta\zeta}(\eta_\zeta)^2$, respectively. Thus, in order to compute a uniformly valid periodic solution, we make use of the Poincar\'{e}-Lindstedt method \cite{29}. We $\zeta = \omega \xi$ and expand the $\omega$ as $\omega= 1 + \delta\,\omega_1 + \delta^2\,\omega_2 + \ldots$
in Eq. (\ref{Reduced}) to obtain,
$$
[1 + 4 \omega_1 \delta + (6 \omega_1^2 + 4 \omega_2)\delta^2 + \ldots] \eta_{\zeta\zeta\zeta\zeta} +
2 {\hat \mu}^2[ 1 + \delta \omega_1 + \delta^2 (\omega_1^2 + 2 \omega_2) + \ldots] \eta_{\zeta\zeta}
$$
$$
+ ({\hat \mu}^2 - \delta \beta_1 \delta^2 \beta_2) \eta=
[1 + 4 \omega_1 \delta + (6 \omega_1^2 + 4 \omega_2)\delta^2 + \ldots] \bigl(\eta_{\zeta\zeta}(\eta_\zeta)^2 +
$$
\begin{equation}
\label{Lind}
 \Gamma (\eta_\zeta \eta_{\zeta\zeta\zeta} + (\eta_{\zeta\zeta})^2)\bigr).
\end{equation}
At leading order we find 
$$
{\cal L}(\eta^{(0)}) {\buildrel \rm def \over =} \eta_{\zeta\zeta\zeta\zeta}^{(0)} + 2 {\hat \mu}^2 {\eta_{\zeta\zeta}}^{(0)} + {\hat \mu}^4 \eta^{(0)}=0
$$
whose normalized solution is given by $\eta^{(0)} = \sqrt{2} \cos{({\hat \mu} \zeta)}$. At $O(\delta^2)$, we have
\begin{equation}
\label{ord2}
{\cal L}(\eta^{(1)}) = \sqrt{2} \beta_1 \cos{({\hat \mu} \zeta)} + 2 {\hat \mu}^4\,\Gamma \cos{(2 {\hat \mu} \zeta)}.
\end{equation}
We set $\beta_1=0$ to remove any mixed-secular terms and solve the resulting equation to find
\begin{equation}
\label{eta2}
{\cal L}(\eta^{(0)}) {\buildrel \rm def \over =}
\eta^{(1)} = (2 \Gamma/9)\,\cos{(2 {\hat \mu} \zeta)}.
\end{equation}
The requirement $\beta_1=0$ translates into the non-existence of subcritical instability.
Proceeding to the $O(\delta^3)$ problem, we have
$$
{\cal L}(\eta^{(2)}) = \sqrt{2}{\hat \mu}^4\,[-5 \omega_1^2 + (\beta_2/{\hat \mu}^4)-1/2 - \Gamma/9]\cos{({\hat \mu}\zeta)}+
(8 \omega_1 {\hat \mu}^4 \Gamma/3) \cos{(2 {\hat \mu} \zeta)}+
$$
\begin{equation}
\label{ord3}
\sqrt{2}{\hat \mu}^4(\Gamma - 1/2)\,\cos{(3 {\hat \mu}\zeta)}
\end{equation}
We set $(-4 \omega_1^2 + (\beta_2/{\hat \mu}^4)-1/2 - \Gamma/9) =0$ and solve for $\omega_1$ value required to remove the secular term,
namely
\begin{equation}
\label{omega1}
\omega_1 = \pm \sqrt{ {\beta_2 \over 4 {\hat \mu}^4} - {4 \Gamma^2+9 \over 72} }
\end{equation}
and upon solving the resulting problem, we find
\begin{equation}
\eta^{(2)} = {-8 \omega_1 \Gamma \over 27} \cos{(2 {\hat \mu} \zeta)} + {9 \sqrt{2} +4(\Gamma^2+4 \Gamma) \over 1152} \cos{(3 {\hat \mu} \zeta)}.
\end{equation}
Thus, a uniformly valid steady state solution to Eq. (\ref{Reduced})  is given by
$$
\eta = \delta\,\sqrt{2} \cos{( {\hat \mu}(1 + 2 \delta \omega_1)\xi)} + {\delta^2\Gamma \over 9}\,\cos{( 2{\hat \mu}(1 + \delta \omega_1)\xi)}+ 
$$
\begin{equation}
\label{uniform}
\delta^3\, [{8 \omega_1 \Gamma \over 27} \cos{( 2{\hat \mu}(1 + \delta \omega_1)\xi)} +
{9 \sqrt{2} +4(\Gamma^2+4 \Gamma) \over 1152}]\,\cos{3( {\hat \mu}(1 + \delta \omega_1)\xi)}  
\end{equation}
Equation (\ref{omega1}) puts a restriction on the validity of Eq.(\ref{uniform}) in the sense that for any deviation from the linear instability threshold measured by $\beta_2$, since $\beta_1=0$, there exists a critical length for the periodic box, $\lambda_C$, below which a solution does not exist, namely
\begin{equation}
\label{condition}
\lambda_C = 2 \pi \bigl( {4 \Gamma^2+9 \over 18 \beta_2 } \bigr)^{1/4}
\end{equation}
The leading order contribution to the velocity, $\phi_0$, is plotted in Fig. (\ref{fig:11}) for $\beta_2$ values ranging from $1$ to $12$ in the periodic box $(-\pi/{\hat \mu}, \pi/{\hat \mu})$ where ${\hat \mu}$ is arbitrarily set equal to $0.65$. We observe the following pattern in the dependence of $\phi_0$ on the parameter ${\hat \beta}$. Small deviations of $\beta$ from $1$ favor the formation of a large centrally located positive cell that is surrounded by a pair of thinner counter-rotating cells, while larger deviations lead to the widening of the pair of the counter-rotating cells accompanied by the squeezing of the positive cell as shown in Fig. (~\ref{fig:11}).
Because the leading order concentration is independent of $z$, its plot depicts a periodic array of descending and rising plumes. At the order $\delta$ contribution, $c_2$, however, also depicted in Fig. (~\ref{fig:11}) show a more involved mechanism of self-organization
for the exchange between the top and lower layers that takes place at the interface. While the descending carbon-rich plumes seem to form an inverted "Y" pattern, the rising carbon-free plumes seem to form a "Y" pattern as shown schematically in Fig. (~\ref{fig:12}).

\section{Discussion and concluding remarks}

Many theoretical studies have attempted to model the convective stability problem that arise in the study of geologic carbon sequestration. Most of the studies have adopted the boundary layer instability approach pioneered by Foster \cite{12} which, due to the time dependence of the base concentration profile, yields instability threshold conditions in terms of critical time. The latter is defined as the time it takes for convection to manifest. In this paper, we have modeled this convection problem by assuming that the leaked carbon dioxide accumulates at the top and forms a layer of carbon saturated brine that overlies a carbon-free brine layer. The resulting step function stratification remains stable until the thickness, and likewise the density, of the carbon saturated layer is sufficient to induce the fluid motion. The model considers a base state consisting a lighter brine-free layer of thickness $(Z_0\,d)$ that underlies a heavier carbon saturated brine of concentration $C_0$ of thickness $d(1-Z_0)$ with a jump in density at $(Z_0\,d)$. This situation is reminiscent of the Rayleigh-Taylor problem wherein a heavy layer sits on top of a lighter layer and separated by a free interface. In our model, the horizontal interface separating the two layers is not free to deform and it allows for buoyancy diffusion to take place \cite{8}. We conducted a linear stability analysis to determine the minimum thickness of the layer of saturated brine  that is required for the manifestation of fluid motion. 

We have considered a Rayleigh-B\'{e}nard set-up with an impervious lower boundary and two types of conditions for the concentration of brine at the upper boundary. The instability threshold parameters consisting of the Rayleigh number, $R$, and associated wavenumber, 
$\alpha$, depend on $Z_0$ as shown in Figs. (\ref{fig:2}, \ref{fig:3}, \ref{fig:6}, and \ref{fig:10}). Various parameters such as thermal diffusivity ratio, reaction rate, and permeability are varied in this study. Tables 1,2, and 3 show the values of the minimum Rayleigh number, critical wavenumber, and critical depth for the case of carbon dioxide concentration being maintained constant at the upper boundary.

Thus, convection sets in when the thickness of the carbon-rich boundary layer exceeds that of $Z_c$ the total thickness of fluid layer.
The critical depth of instability is found to depend on the imposed boundary conditions but not on the depth of the sequestration site. For instance, for symmetric boundary conditions at the top and bottom boundaries, the critical depth equal half that of the total depth of the layer. 
The associated streamlines, shown in Fig. (\ref{fig:5}), have the typical oval shape except that the ellipses are centered at the $Z_0$ level and not at the mid-plane of the fluid layer. 
The convective iso-concentration contours, however, are uncommon.
While the contours have the typical oval shape or lens-shape above the $Z_0$ interface, they exhibit an unusual tongue-like shape below $Z_0$. The center-most contours have the shape of a round tongue which evolve to shapes that are first triangular and then elongated acute triangular farther away from the center. Furthermore, the contours are open at the lower boundary and continuous but not smooth where they connect at $Z_0$. This jump in the first derivative is quantified by
$$
	DS^+-DS^-=-p S^-(\tanh(p Z_0)+\coth(p(1-Z_0))
$$
 When the carbon dioxide is allowed to leak into the brine solution by assuming an upper boundary that is nearly-impervious, we find the minimum Rayleigh number and associated wavenumber to be $R_C=8$ and $\alpha_C=0$, respectively which are attained at half the depth the fluid layer.

The weakly nonlinear analysis was carried out using long wavelength asymptotics for the nearly impermeable boundary. We derived an evolution equation for the leading order concentration $f$ Eq. (\ref{uniform}) and in the process of the derivation, we obtained an analytical expression for $R_C$, Eq. (\ref{RC}), which agrees with the numerical results. We find that the leading order concentration $c_0=f$ is smooth at $Z_0$ and that the singular behavior of $c$ appears at $O(\epsilon^2)$ where, from Eq. (\ref{DC2}), we note that
$c$ has a jump in the first derivative  of $c_2$ at $Z_0$ given by
$$
{Dc_2}^+(Z_0)-{Dc_2}^-(Z_0) = f_{XX} = - c_2(Z_0)
$$
This jump, which is independent of the location of $Z_0$, indicates that at the order $\delta^2$, the carbon concentration satisfies a Newton law of cooling condition.
We investigated the stability of the trivial solution, $\eta=0$, which corresponds to the base state and derived the dispersion relation, Eq. (\ref{Disp2}), to determine the instability condition, ${\hat \mu}^4 > {\hat \beta}$ and critical wavenumber $\gamma={\hat \mu}$.
We also investigated the effects of the nonlinear terms in the evolution equation by a perturbation analysis around the linear solution
by making use of the Lindstedt-Poincar\'{e} method to obtain a uniformly valid super-critical solution, Eq. ({\ref{uniform}). Furthermore, the analysis also unearthed a condition, Eq. (\ref{condition}), that must be satisfied for super-critical 
periodic solutions to exist. In summary, the configuration for the base state considered in this paper leads to stability characteristics and flow features that are qualitatively similar to those observed in experiments. Direct comparison cannot be made given that the experiments are carried  out in finite enclosures and at higher Rayleigh numbers (see \cite{13}-\cite{14}).

	\section{Acknowledgements}
	We thank Liet Vo for useful conversations and two anonymous reviewers whose comments have improved the presentation of the manuscript.

	\newpage
	
	\section{Appendix}
	
	In the process of conducting the linear stability analysis, the homogeneous linear system of equations obtained through the matching conditions at $Z_0$ is described by 
\begin{equation}
\label{18}
	\left\{
	\begin{array}{ll}
	Q_{11}\,A^- + Q_{12}\,B^-+ Q_{13}\,A^+ + Q_{14}\,B^+ &= 0\nonumber \\
	Q_{21}\,A^- + Q_{22}\,B^-+ Q_{23}\,A^+ + Q_{24}\,D^+ &= 0\nonumber \\
	Q_{31}\,A^- + Q_{32}\,B^-+ Q_{33}\,A^+ + Q_{34}\,B^+ &= 0\nonumber \\
	Q_{41}\,A^- + Q_{42}\,B^-+ Q_{43}\,A^+ + Q_{44}\,B^+ &= 0\nonumber \\
	\end{array}
\right.
\end{equation}
	where\\
	$Q_{11}=\sinh{(\alpha Z_0)}$, $Q_{12}=Z_0\,\sinh{(\alpha Z_0)} +(\beta/\alpha) Z_0\,\cosh{(\alpha Z_0)}$, $Q_{13}=-\sinh{(\alpha (Z_0-1))}$,\\
	$Q_{14}=-(Z_0-1)\,\sinh{(\alpha (Z_0-1))} +(\beta/\alpha) (Z_0-1)\,\cosh{(\alpha (Z_0-1))}$\\
	$Q_{21}=\alpha\,\cosh{(\alpha Z_0)}$,  $Q_{22}=(1+\beta\,Z_0)\,\sinh{(\alpha Z_0)}+(\alpha Z_0-\beta/\alpha)\,\cosh{(\alpha Z_0)}$\\
	$Q_{23}=-\alpha \cosh{(\alpha (Z_0-1))}$, \\
	$Q_{24} = -(1-\beta (Z_0-1))\sinh{(\alpha (Z_0-1)) }- (\alpha (Z_0-1)-\beta/\alpha)\cosh{(\alpha(Z_0-1)}$\\
	$Q_{31} = \alpha^2 \sinh{(\alpha Z_0)}$, $Q_{33}= -\alpha^2 \sinh{(\alpha (Z_0-1))}$\\
	$Q_{32} = (2\alpha+\beta \alpha Z_0)\cosh{(\alpha Z_0)} + (\alpha^2 Z_0 + 2 \beta)\sinh{(\alpha Z_0)}$\\
	$Q_{34} = -(2\alpha - \beta \alpha (Z_0-1)) \cosh{(\alpha (Z_0-1))} - (\alpha^2 (Z_0-1)-2 \beta) \sinh{(\alpha (Z_0-1))}$ \\
	$Q_{41} = -\alpha^3\,\cosh{(\alpha Z_0)}- R \alpha^2 \sinh{(\alpha Z_0)}$, \\
	$Q_{43} = \alpha^3\,\cosh{(\alpha (Z_0-1))}$\\
	$Q_{42} = -(3 \alpha\,\beta + Z_0\,\alpha^3) \cosh{(\alpha Z_0)} - (3 \alpha^2 + \beta Z_0 \alpha^2) \sinh{(\alpha Z_0)}
	- R \alpha^2 (Z_0 \sinh{(\alpha Z_0)} + Z_0 (\beta/\alpha) \cosh{(\alpha (Z_0-1)})$\\
	$Q_{44} = -(3 \alpha\,\beta + (Z_0-1) \alpha^3) \cosh{(\alpha (Z_0-1)} + (3 \alpha^2 - \beta (Z_0-1) \alpha^2) \sinh{(\alpha(Z_0-1))}$ \\
	$P_{41}=-\alpha^3\,\cosh{(\alpha Z_0)}$, $P_{42}=-(3 \alpha\,\beta + Z_0\,\alpha^3) \cosh{(\alpha Z_0)} - (3 \alpha^2 + \beta Z_0 \alpha^2) \sinh{(\alpha Z_0)}$\\
	$M_{41} = \alpha^2 \sinh{(\alpha Z_0)}$, $M_{42}=\alpha^2 (Z_0 \sinh{(\alpha Z_0)} + Z_0 (\beta/\alpha) \cosh{(\alpha (Z_0-1)})$,\\
	
\noindent
The matrices ${\bf Q}_1$ and ${\bf Q}_2$ are given by,

	\begin{equation*}
	{\bf Q_1}=\left[ 
	\begin{array}{cccc}
	Q_{11} & Q_{12} & Q_{13} & Q_{14} \\ 
	Q _{21} & Q_{22} & Q_{23} & Q_{24} \\ 
	Q_{31} & Q_{32} & Q_{33} & Q_{34} \\ 
	P_{41} & P_{42} & Q_{43} & Q_{44}
	\end{array}
	\right],
	\qquad 
	{\bf Q_2}=\left[ 
	\begin{array}{cccc}
	Q_{11} & Q_{12} & Q_{13}  & Q_{14} \\ 
	Q _{21} & Q_{22} & Q_{23} & Q_{24} \\ 
	Q_{31} & Q_{32} & Q_{33} & Q_{34} \\ 
	M_{41} & M_{42} & 0 & 0
	\end{array}
	\right].
	\quad 
	\end{equation*}
	
	\vspace{0.2in}
	
	The eigenfunctions for the linear problem for the case $c=0$ at $z=1$ are given by
	\begin{eqnarray}
	W^-(z)&=&(1-1.3783\,z)\sinh{(\alpha_C z)}\\
	W^+(z)&=& 0.9482(z-1)\,\cosh{(\alpha_C(z-1))}-0.8876 \sinh{(\alpha_C(z-1))}
	\end{eqnarray}
	
	$$
	{\cal J}= \frac {\alpha_C^2\,(1-1.3783 Z_{0C}) \sinh{(\alpha_C Z_{0C})}}
	{\alpha_C\,\cosh{\alpha_C}\bigl(\cosh{(\alpha_C(1-Z_{0C})}/\sinh{(\alpha_C(1-Z_{0C})} \bigr)-1}
	$$
	\begin{eqnarray}
	S^-(z)&=&-{\cal J} \cosh{(\alpha_C z)}\\
	S^+(z)& =& \frac {{\cal J} \cosh{(\alpha_C Z_{0C})} }{\sinh{(\alpha_C(1-Z_{0C}))}}\,\sinh{(\alpha_C(1-z))}
	\end{eqnarray}
	where $\alpha_C=2.4$ and $Z_{0C}=0.38$.


\newpage
	
	
	
	
	
	
	
	
	
	
	\begin{thebibliography}{99}
		
		\bibitem{1} Change, Intergovernmental Panel On Climate PCC (2007) Aspectos Regionais e Setoriais da Contribuio
		do Grupo de Trabalho II ao 4 Relatrio de Avaliao Mudana Climtica 2007 do IPCC.
		
		\bibitem{2} Plain CO2 Reduction (PCOR) Partnership. http://www.undeerc.org/pcor/
		
		\bibitem{3} Pacala S, Socolow R (2004)Wedges: Solving the Climate Problem for the Next 50 Years with Current
		Technologies. Science 305: 968-972.
		
		\bibitem{4} Matter JM, Stute M, Snbjrnsdottir SO, Oelkers EH, Gislason SR, Aradottir ES, Sigfusson B, Gunnarsson
		I, Sigurdardottir H, Gunnlaugsson E, Axelsson G, Alfredsson HA,Wolff-Boenisch D, Mesfin
		K, Diana Fernandez de la Reguera Taya, Hall J, Knud Dideriksen10, Broecker WS (2016) Rapid carbon
		mineralization for permanent disposal of anthropogenic carbon dioxide emissions. Science 352:
		1312-1314 (DOI: 10.1126/science.aad8132 )
		
		\bibitem{5} Barba Rossa G, Cliffe KA, Power H (2016). Effects of hydrodynamic dispersion 
		on the stability of
		buoyancy-driven porous media convection in the presence of first order chemical reaction. J. Eng.
		Math. (DOI 10.1007/s10665-016-9860-z).
		
		\bibitem{6} De Paoli M, Zonta F, Alfredo Soldati A (2016) Influence of anisotropic permeability on convection in
		porous media: Implications for geological CO2 sequestration. Phys. Fluids 28:056601
		
		\bibitem{7} Xu X, Chen S, Zhang D (2006) Convective stability analysis of the long term storage of carbon dioxide
		in deep saline aquifers. Advances in Water Resources 29: 397-407.
		
		\bibitem{8} Batchelor GK, Nitsche J (1991) Instability of stationary unbounded stratified fluid. J. Fluid Mech.
		227: 357-391.
		
		\bibitem{9} Simitev RD, Busse FH (2010) Problems of astrophysical convection: thermal convection in layers
		without boundaries. Proceedings of the 2010 Summer Program. Center of Turbulence Research, Stanford University, 485-492.
		
		\bibitem{10} Hadji L, Shahmurov S, Aljahdaly NH (2016) Thermal convection induced by an infinitesimally thin
		and unstably stratified layer. J. Non-Equilib. Themodyn. (DOI 10.1515/jnet-2015-0071).
		
		\bibitem{11} Foster T (1965) Onset of convection in a layer of fluid cooled from above. Phys. Fluids 8: 1770-1774		
		
		\bibitem{12} Foster T (1969) Onset of manifest convection in a layer of fluid with a time-dependent surface temperature.
		Phys. Fluids 12: 2482-2487.
		
		\bibitem{13} Kneafsey, TJ,  Karsten P (2010) Laboratory flow experiments for visualizing carbon dioxide-induced,  
		density-	driven 	brine convection. Transport in porous media 82: 123-139.
		
		\bibitem{14} Neufeld, JA, Hesse MA, Riaz A, Hallaworth MA, Tchelepi HA (2010)  Convective dissolution of carbon dioxide in 		saline aquifers. Geophys. Res. Lett.  37: L22404	
		
         \bibitem{15} Hill AA, Morad MR (2014) Convective stability of carbon sequestration in anisotropic porous media.
		Proc. R. Soc. A 470: 20140373.
		
		\bibitem{16} Bhadauria BS (2012) Double-diffusive convection in a saturated anisotropic porous layer with internal
		heat source. Transport in Porous Med. 92: 299-320.	
		
		\bibitem{17} Horton CW, Rogers Jr FT (1945) Convection currents in a porous media. J. Appl. Phys. 16: 367-370.	
			
		\bibitem{18} Lapwood ER (1948) Convection of a fluid in a porous medium. Proc. Phil. Soc. 44: 508-521.
			
		\bibitem{19} De La Torre Juarez M, Busse FH (1995) Stability of two-dimensional convection in a fluid saturated
		porous medium. J. Fluid Mech. 292: 305-323.
		
		\bibitem{20} Barletta A, Tyvand PA, Nyg ˙ ard (2015) Onset of thermal convection in a porous layer with mixed
		boundary conditions. J. Eng. Math. 91:105-120.		
		
		\bibitem{21} Riaz A, Hesse M, Tchelepi HA, Orr FM (2006) Onset of convection in a gravitationally unstable diffusive boundary 
		 layer in porous media. J. Fluid Mech. 548: 87-111
		 
		 \bibitem{22} Ennis-King J, Preston I, Paterson L (2005) Onset of convection in anisotropic porous media subject to a rapid  			change in boundary conditions. Phys. Fluids 17: 084107
				
		\bibitem{23} Slim AC, Ramakrishnan TS (2010)
		Onset and cessation of time-dependent, dissolution-driven convection in porous media
         Phys. Fluids 22: 124103. 
		
		\bibitem{24} Slim AC (2014) Solutal-convection regimes in a two-dimensional porous medium. J. Fluid Mech. 741: 461-491
		
		\bibitem{25} Klinkenberg K (1941) The permeability of porous media to liquids and gases. Drilling and production
		practice. American Petroleum Institute.
		
		\bibitem{26} Rasenat, S, Busse FH, Rehberg I (1989) A theoretical and experimental study of double-layer convection.
		J. Fluid Mech. 199: 519-540.
		
		\bibitem{27} Busse FH, Riahi N (1980) Nonlinear thermal convection with poorly conducting boundaries, J. Fluid
		Mech. 96: 243-256
		
		\bibitem{28} Chapman CJ, Proctor MRE 1980 Nonlinear Rayleigh-B\'{e}nard convection between poorly conducting
		boundaries. J. Fluid Mech. 101: 759-782.
		
		\bibitem{29} Nayfeh AH (2004) Introduction to Perturbation Techniques, Wiley-VCH, 139.
		
		
		
		
		
	\end{thebibliography}
\end{document}